\documentclass[, a4paper,twocolumn,aps,prx,longbibliography,superscriptaddress]{revtex4-2}
 
\usepackage{amsmath}
\usepackage{pifont}
\usepackage{amssymb}
\usepackage{amsthm}
\usepackage{graphicx}
\usepackage{xcolor}
\usepackage[english]{babel}
\usepackage{csquotes}
\usepackage{braket}
\usepackage{hyperref}
\usepackage{mathtools}
\usepackage{enumitem}
\usepackage{IEEEtrantools}
\usepackage{relsize}
\usepackage{placeins}
\usepackage{comment}
\usepackage[normalem]{ulem}
\usepackage{systeme}
\usepackage{physics}
\usepackage{algpseudocode,algorithm}
\usepackage{times}
\usepackage{booktabs}
\usepackage{titlesec}

\usepackage{algorithm}
\usepackage{algpseudocode}

\usepackage{bm}
\usepackage{tikz-cd} 
\usepackage{adjustbox}

\usepackage{tikz}						
\usetikzlibrary{math, arrows,shapes.misc,snakes,
		       automata,backgrounds,
		       petri,topaths, decorations.pathmorphing, tikzmark}	

\usepackage{pgffor}

\usepackage{pgfplots}
\usetikzlibrary{pgfplots.groupplots}
\pgfplotsset{compat=1.11}
\usepgfplotslibrary{fillbetween}

\usepackage{tikz}
\usetikzlibrary{
  shapes,
  shapes.geometric,
	trees,
	matrix,
  positioning,
    pgfplots.groupplots,
  }

\PassOptionsToPackage{pdftex,hyperfootnotes=false,pdfpagelabels}{hyperref}
\hypersetup{
	colorlinks = true,
	linkcolor = [rgb]{0.70,0.13,0.13},
	citecolor = [rgb]{0.13,0.55,0.13},
	urlcolor = [rgb]{0.25, 0.41, 0.88}}
\usepackage[capitalize]{cleveref}

\newtheorem{theorem}{Theorem}

\newtheorem{observation}[theorem]{Observation}

\begin{document}

\title{Minimizing classical resources in variational measurement-based quantum computation for generative modeling}

\author{Arunava Majumder}
\affiliation{University of Innsbruck, Department of
Theoretical Physics, Technikerstraße 21a, A-6020 Innsbruck, Austria.}
\author{Hendrik Poulsen Nautrup}
\affiliation{University of Innsbruck, Department of
Theoretical Physics, Technikerstraße 21a, A-6020 Innsbruck, Austria.}
\author{Hans J. Briegel}
\affiliation{University of Innsbruck, Department of
Theoretical Physics, Technikerstraße 21a, A-6020 Innsbruck, Austria.}

\begin{abstract}
Measurement-based quantum computation (MBQC) is a framework for quantum information processing in which a computational task is carried out through one-qubit measurements on a highly entangled resource state. Due to the indeterminacy of the outcomes of a quantum measurement, the random outcomes of these operations, if not corrected, yield a variational quantum channel family. Traditionally, this randomness is corrected through classical processing in order to ensure deterministic unitary computations. Recently, variational measurement-based quantum computation (VMBQC) has been introduced to exploit this measurement-induced randomness to gain an advantage in generative modeling. A limitation of this approach is that the corresponding channel model has twice as many parameters compared to the unitary model, scaling as $N \times D$, where $N$ is the number of logical qubits (width) and $D$ is the depth of the VMBQC model. This can often make optimization more difficult and may lead to poorly trainable models. In this paper, we present a restricted VMBQC model that extends the unitary setting to a channel-based one using only a single additional trainable parameter. We show, both numerically and algebraically, that this minimal extension is sufficient to generate probability distributions that cannot be represented by the corresponding unitary model. 
\end{abstract}

\maketitle

\section{Introduction}
Quantum computers are expected to outperform classical computers for certain classes of problems~\cite{nielsen2010quantum, arute2019quantum, daley2022practical, google2025observation, shor1999polynomial}. In the standard circuit model, quantum computation proceeds deterministically through a sequence of unitary gates, followed by measurements performed only at the final stage to extract the output. Measurement-based (or one-way) quantum computation (MBQC)~\cite{briegel2009m,raussendorf2001a,Raussendorf2003cluster} offers a fundamentally different approach to perform computation. In this paradigm, one first prepares a highly entangled resource state, e.g., a cluster state~\cite{briegel2001persistent}, and then computation is carried out entirely via a sequence of one-qubit measurements on this state.

Since measurement outcomes in MBQC are, in general, intrinsically random, the computation naturally gives rise to probabilistic evolution corresponding to a mixed quantum channel. However, by employing fully adaptive measurement schemes that correct unintended outcomes at every step, this randomness can be compensated, effectively reducing the computation to a deterministic unitary evolution. Recent works have shown that the view of MBQC as a quantum channel can be exploited to achieve advantages over both classical approaches~\cite{huang2025generative} and purely unitary quantum models~\cite{majumder2024variational}. In a related direction, Ref.~\cite{coyle2025training} studies density quantum neural networks, randomized mixtures of parametrized unitaries that effectively implement a quantum channel, and demonstrated benefits compared to purely unitary models. Another recent work~\cite{kurkin2025note} shows that parameterized Instantaneous Quantum Polynomial (IQP) circuits with hidden (traced-out) qubits, effectively forming a quantum channel on the visible qubits, are universal for generating arbitrary probability distributions on bit strings. While IQP circuits on $N$ qubits alone are not universal, it was shown that IQP circuits on $2N+1$ qubits with $N+1$ qubits traced out become universal on the probability simplex of the remaining $N$ output bits. This illustrates the effective advantage of channel models over purely unitary models, as a channel implemented on $N$ qubits can realize distributions that a unitary model acting only on $N$ qubits cannot. In particular, Ref.~\cite{majumder2024variational} introduces a quantum learning model based on MBQC embedded in the quantum circuit Born machines (QCBM)~\cite{liu2018d,benedetti2019generative} framework, called variational MBQC (VMBQC), which allows for \textit{controllable} partial correction of measurement outcomes by introducing additional learning parameters that tune the likelihood of correcting the unintended measurement outcomes. This controlled partial adaptation of measurement outcomes gives rise to a trainable quantum channel model that can be shown to outperform its unitary counterpart (where all measurements are corrected), given the same (\textit{quantum}) resources, both analytically and numerically. The authors demonstrate this advantage in a generative learning setting~\cite{Tomczak2022}, where the goal is to learn and generate new samples from a target probability distribution given a limited number of training samples.

While the VMBQC-based channel model does not require additional \textit{quantum} resources, it does require additional classical parameters, namely the correction probabilities, which control the partial adaptation of measurement outcomes. Given a unitary MBQC implemented on a cluster state of width $N$ and depth $D$, the corresponding channel model introduces $N \times D$ additional trainable classical parameters, which gives twice as many as in the unitary model. While increasing the number of parameters may enhance the expressivity of a quantum learning model, it can significantly complicate the classical optimization process~\cite{mcclean2018barren}. In addition, the randomness of measurement outcomes can be interpreted as a form of noise, which may add to a flat loss landscape~\cite{wang2021noise}. This motivates the following question:

\textit{How many additional classical resources are sufficient to generate distributions with the VMBQC-based channel models that cannot be generated by the unitary models?}

Here, classical resources refer to the number of additional trainable parameters introduced to the VMBQC model along with their gradient estimation overhead. In this paper, we show both analytically and numerically that introducing just a single additional trainable parameter is sufficient for the variational quantum channel model to outperform its purely unitary counterpart in generative learning tasks.

This manuscript is structured as follows. In Sec.~\ref{sec:qgm}, we introduce quantum generative modeling and the QCBM framework, while Sec.~\ref{sec:vmbqc_pqc} describes the MBQC framework and the VMBQC model. Sec.~\ref{sec:models} presents detailed descriptions of all training models used in this work. The results are presented in Sec.~\ref{sec:res}. We first show algebraically that there exists a VMBQC channel model with a single additional trainable parameter whose output distribution cannot be represented by the corresponding unitary model. Motivated by this result, we then numerically compare the performances of several channel models, each with a single additional parameter, against the corresponding unitary model.

\section{Methods}\label{methods}
This section reviews classical and quantum generative models (Sec.~\ref{sec:qgm}). Sec.~\ref{sec:vmbqc_pqc} summarizes variational measurement-based quantum computation from Ref.~\cite{majumder2024variational}. Sec.~\ref{sec:models} introduces the learning models used in our analysis, and Sec.~\ref{sec:loss} defines the loss and corresponding gradients.

\subsection{Quantum Generative Modeling }\label{sec:qgm}
\begin{figure*}[t]
\centering
\includegraphics[width=0.95\textwidth]{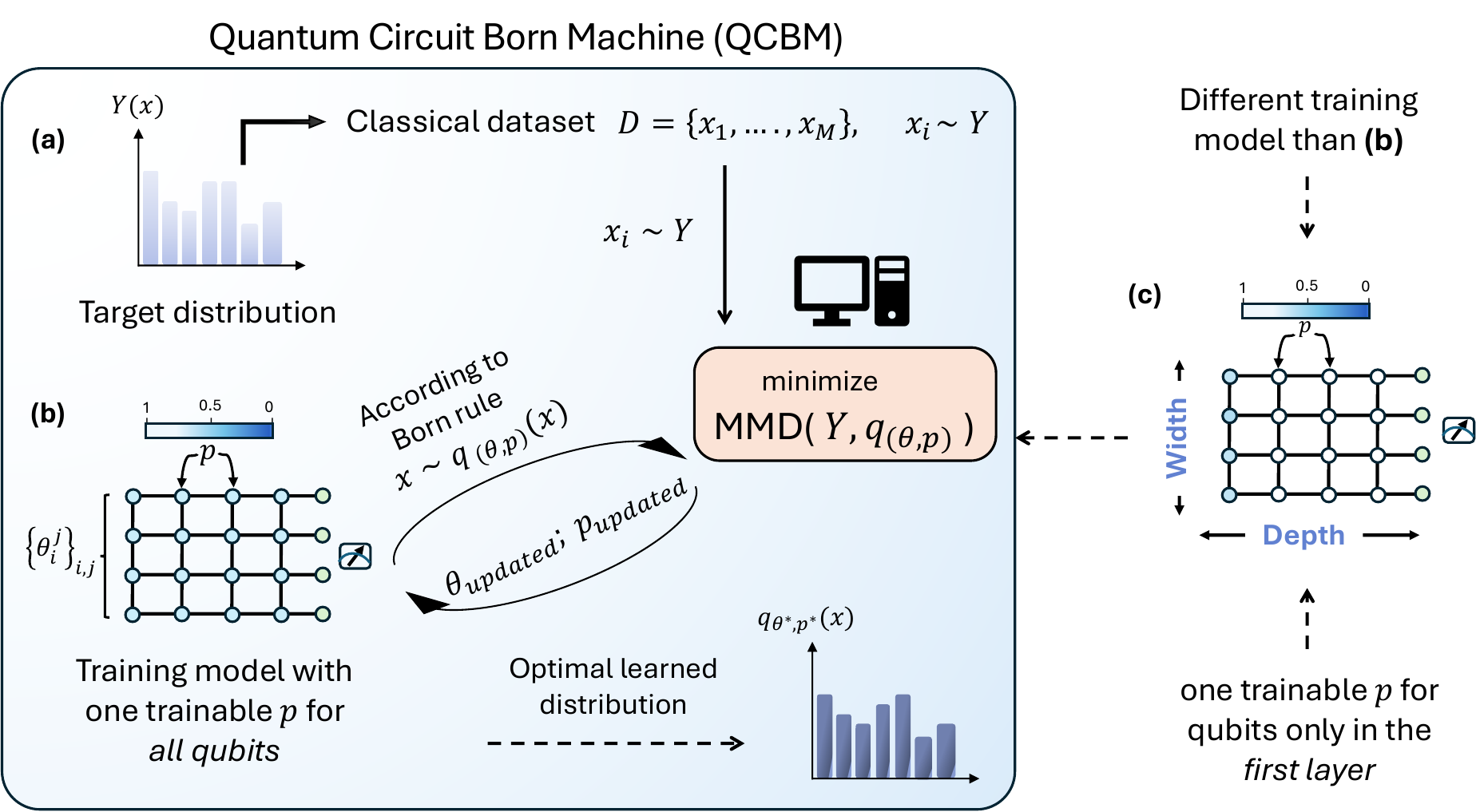}
\caption{\small{\textbf{Quantum circuit Born machines with VMBQC:}} 
\textbf{(a)} Target distribution $Y(x)$ (usually unknown), from which samples are given, $x\sim Y$. In this paper, the target distribution is generated from the VMBQC model itself with a randomly chosen set of parameters $\{\bm{\theta}, \bm{p}\}$ on the cluster state with some given width ($N$) and depth ($D$). \textbf{(b)} The learning model, with width $N$ and depth $D$, is parametrized by measurement angles $\bm{\theta}:=\{\theta^j_i\}_{i,j}$ and \textit{a single correction probability $p$}, that is, $\bm{p}:=\{p^j_i=p\}_{i,j}$, for all partially adapted qubits (\textit{blue}), by which we mean qubits, on which the measurement outcomes are adapted with probability $p$. When measured on the last layer, the model produces bit strings from the underlying distribution ($x\sim q_{(\bm{\theta},\bm{p})}$) according to the Born rule. The MMD loss takes samples from both the learning model and the target distribution and measures the distance between the two distributions $Y$ and $q_{(\bm{\theta},\bm{p})}$, and updates the parameters using the parameter shift rule, on a classical computer. The model receives the updated parameters, and this loop continues until the MMD converges to some minimum value. Once converged, the optimal parameters $\bm{\theta}^{*}$ and $\bm{p}^{*}$ generate the optimal learned distribution. The entire learning process can also be implemented using a different learning model \textbf{(c)} where we allow partially uncorrected qubits ($p^j_i<1$) only on the first layer of the cluster state (\textit{blue}), with the same $p^j_i=p<1$ for all of them, and the training procedure remains identical. All other qubits, in \textit{white}, have $p^j_i=1$, i.e., measurements on those qubits are always corrected. All the models in this study have periodic boundary conditions, i.e., the first and the last qubits in the cluster state are connected. The qubits in \textit{green} at the end of the cluster are the output qubits measured in the computational basis. The samples $D=\{x_1,\cdots,x_M\}$ are used only to evaluate the loss, while the VMBQC model starts from a fixed quantum state.} 
\label{fig:gen_l}
\end{figure*}

Generative learning refers to a class of unsupervised learning methods that aim to model the probability distribution underlying a given data set (assumed to be the output of some probabilistic process). Generative models can be broadly categorized as either \textit{explicit}, which provide a tractable mathematical expression for the distribution they generate, denoted by $q_{(\bm{\theta},\bm{p})}$, or \textit{implicit}, in which case they describe only the stochastic process of sampling from the distribution, i.e., $x \sim q_{(\bm{\theta},\bm{p})}$~\cite{mohamed2016learning}, with $(\bm \theta, \bm p)$ being parameters of the learning model. In this work, we focus on implicit generative models. The training data set for a generative model consists of a collection of $M$ independent and identically distributed ($i.i.d.$) samples $D=\{x_1, x_2, \ldots, x_M\}$, where each $x_{k}\in \{ 0,1 \}^{N}$ is a bit string of length $N$, with $ ~k\in[M]:=\{1,\cdots,M\}$, sampled from some unknown target distribution $Y(x)$. The goal is to design and train an implicit model such that its optimal empirical distribution $q_{(\bm{\theta^{*}},\bm{p^{*}})}(x)$ (defined by its optimal set of parameters $\{\bm{\theta^{*}},\bm{p^{*}}\}$) approximates $Y(x)$  within some desired accuracy (see Fig.~\ref{fig:gen_l}). Note that, in the generative machine learning setting considered here, the samples in $D$ are not supplied individually as inputs to the learning model. Rather, they are used to characterize the target distribution and evaluate the training loss. Therefore, no sample-wise classical data encoding is required. The learning model starts from a fixed quantum state.

\vspace{8pt}

Classical generative models typically rely on neural networks with trainable weights to learn a target probability distribution. In quantum generative modeling, for example in quantum circuit Born machines, the neural network is replaced by a parametrized quantum circuit (PQC). Accordingly, the QCBM framework consists of several components: $(i)$ a sufficiently expressive PQC composed of parametrized single-qubit gates and entangling two-qubit gates, $(ii)$ a classical post-processing stage in which a cost function and its corresponding gradients are evaluated using samples generated by the PQC, and $(iii)$ a feedback loop that iteratively updates the PQC parameters until convergence is reached. The QCBM can thus be summarized as follows. First, a parametrized quantum state $\ket{\psi(\bm{\theta}, \bm{p})}$ is prepared by applying the PQC, with trainable parameters $\{\bm{\theta}, \bm{p}\}$, to a fixed initial state. Measurements performed at the end of the circuit, in a chosen basis, then generate samples according to the Born rule $q_{(\bm{\theta}, \bm{p})}(x)=|\langle x|\psi(\bm{\theta}, \bm{p})\rangle|^{2}$. Each execution of the PQC yields a single sample (bitstring $x$) drawn from the model distribution $q_{(\bm{\theta}, \bm{p})}(x)$. By collecting sufficiently many such samples, one can perform loss optimization using an appropriate loss function (see Fig.~\ref{fig:gen_l}). A key advantage of QCBMs is their ability to efficiently generate unbiased samples from the learned distribution, which is a fundamental requirement for generative learning.  The VMBQC framework introduced in Ref.~\cite{majumder2024variational} provides a natural realization of PQCs by leveraging a circuit-model representation of MBQC in terms of so-called Clifford Quantum Cellular Automata (CQCA) (see Ref.~\cite{nautrup2023measurement}). The relevant construction is reviewed in Sec.~\ref{sec:vmbqc_pqc}.

\subsection{MBQC with probabilistic
byproduct correction}\label{sec:vmbqc_pqc}
The discussion below provides a concise summary of the framework developed in Refs.~\cite{nautrup2023measurement, majumder2024variational}, which serve as the foundation for the extension introduced in the present work.

\begin{figure}[t]
\centering
\includegraphics[width=0.47\textwidth]{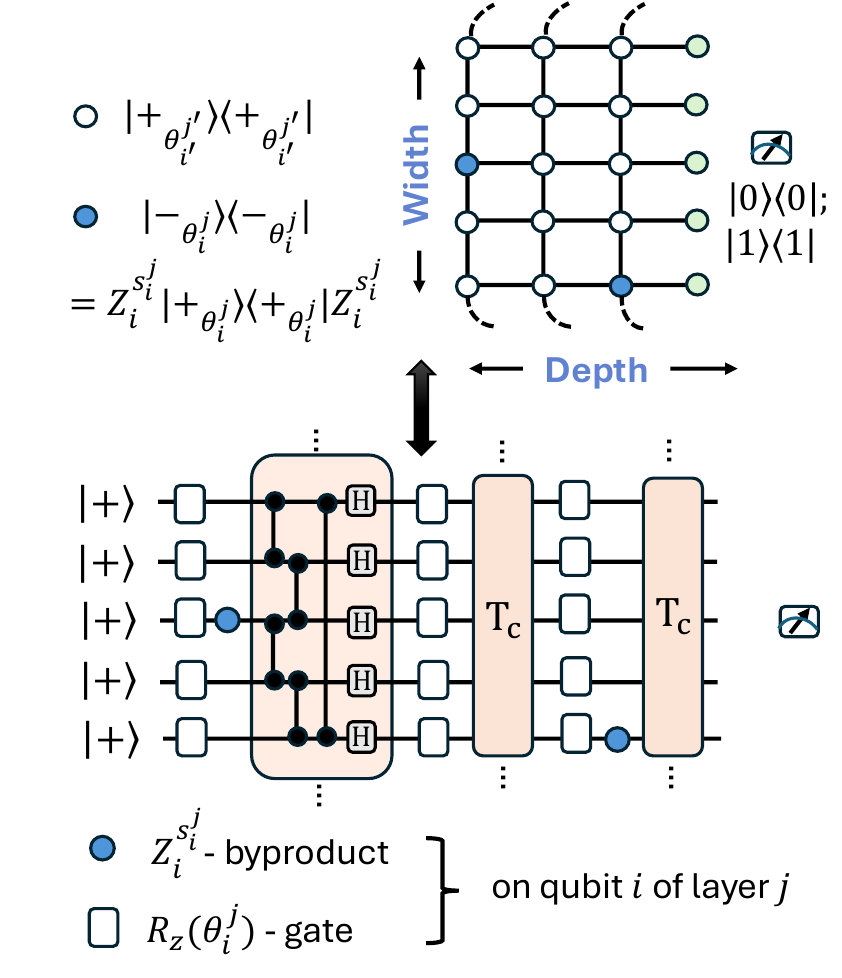}
\caption{\small{\textbf{C-QCA based circuit translation of an MBQC on cluster state:} An MBQC on the cluster state (top), with width $N=5$ and depth $D=3$, can be understood in the circuit picture (bottom) where $XY$-plane measurements become $Z$-rotations in the circuit, interleaved with Clifford blocks $T_c$. The fourth column (top), with output qubits (\textit{green}), is measured in the computational $Z$-basis. An unintended measurement outcome, e.g., when the $(i,j)=(3,1)$ qubit (blue) in the cluster state is measured in the $e^{-i\theta_3^1 Z/2}\ket{-}\bra{-}e^{i\theta_3^1 Z/2}$-basis, yields the outcome $-1$, resulting in an equivalent byproduct operator $Z_i^{s_i^j}$ appearing in the circuit. The qubits in white yield a $+1$ outcome. All models have periodic boundary conditions, i.e., the first(top) and last(bottom) qubits are connected with a $CZ$ gate.} } 
\label{fig:cqca}
\end{figure}

In MBQC, one can perform universal quantum computation using single-qubit measurements on a highly entangled resource state~\cite{raussendorf2001a,briegel2001persistent}. The most common resource state is the so-called cluster state, which can be prepared by placing each qubit in the $\ket{+}$ state on a rectangular lattice and applying entangling $CZ$ gates between neighbors [Fig.~\ref{fig:cqca} (top)]. We always assume periodic boundary conditions on the cluster state, connecting the qubit $(1,j)$ and qubit $(N,j)$ for all $j\in[1,D]$. Computation is carried out by measuring each qubit at site $(i,j);~ i\in [N],j\in[1,D]$ in the cluster state in the $XY$-plane~\cite{Mantri2017universality}, i.e., measuring the qubit in the basis $\ket{\pm_{\theta_i^j}} := 1/\sqrt{2(}\ket{0} \pm e^{-i\theta_i^j}\ket{1})$. Here, $N$ is the width of the cluster and $D$ is its depth. The measurement angles $\theta_i^j$ then determine the effective unitary gate that is applied. The lattice has a total size of $N\times(D+1)$, where the final column (layer with qubits in \textit{green} in Fig.~\ref{fig:cqca}) serves as the output layer and is typically measured in the computational $Z$-basis. 


When measurements are performed in the aforementioned basis on the cluster state, the post-measurement state is projected, with equal probability, onto either $\ket{+_{\theta}}\bra{+{_\theta}}$ or $\ket{-_{\theta}}\bra{-_{\theta}}$. The $+1$ measurement outcome directly implements the intended computation, whereas the $-1$ outcome realizes the same computation up to an additional Pauli-$Z$ operator, since $\ket{-_{\theta}}\bra{-_{\theta}} = Z \ket{+_{\theta}}\bra{+_{\theta}} Z$. These $Z$ operators are referred to as byproduct operators and can be compensated through classical processing and adaptive measurement updates \cite{majumder2024variational}.

MBQC on a cluster state with $XY$-plane measurements is universal for quantum computation (Ref.~\cite{Mantri2017universality}) and admits a well-defined circuit model representation, as established in Refs.~\cite{nautrup2023measurement,stephen2019subsystem,raussendorf2019computationally,raussendorf2005quantum}. As illustrated in Fig.~\ref{fig:cqca} (bottom), each $XY$-plane measurement on qubit $i$ in layer $j$ corresponds to a single qubit $Z$-rotation, $R_z(\theta^j_i)=e^{-i Z_i\theta^j_i/2}$, in the circuit model. A measurement outcome $+1$ (white), equivalently $s_i^j=0$, implements the intended $Z$-rotation without any additional byproduct. A measurement outcome $-1$ (blue), equivalently $s_i^j=1$, introduces a Pauli-$Z_i$ byproduct immediately after the rotation. Both cases are compactly represented by $Z_i^{s_i^j}$ in the Fig.~\ref{fig:cqca}, where $s_i^j\in\{0,1\}$ denotes the binary measurement outcome. These operations are followed by a fixed Clifford layer $T_c$, consisting of nearest-neighbor $CZ$ gates (with periodic boundary conditions) and Hadamard gates applied to all qubits (see Fig.~\ref{fig:cqca}). Further details can be found in Ref.~\cite{nautrup2023measurement,majumder2024variational}.

The fixed Clifford layer associated with one column of the periodic cluster state is \begin{equation} T_c := H^{\otimes N} \left( \prod_{i=1}^{N-1} CZ_{i,i+1} \right) CZ_{N,1}. \end{equation}
along with the rotation convention
\begin{equation} R_{z}(\theta_i^j) := \exp\left(-i\theta_i^j Z_i/2\right). \end{equation}


\begin{figure}[t]
\centering
\includegraphics[width=0.5\textwidth]{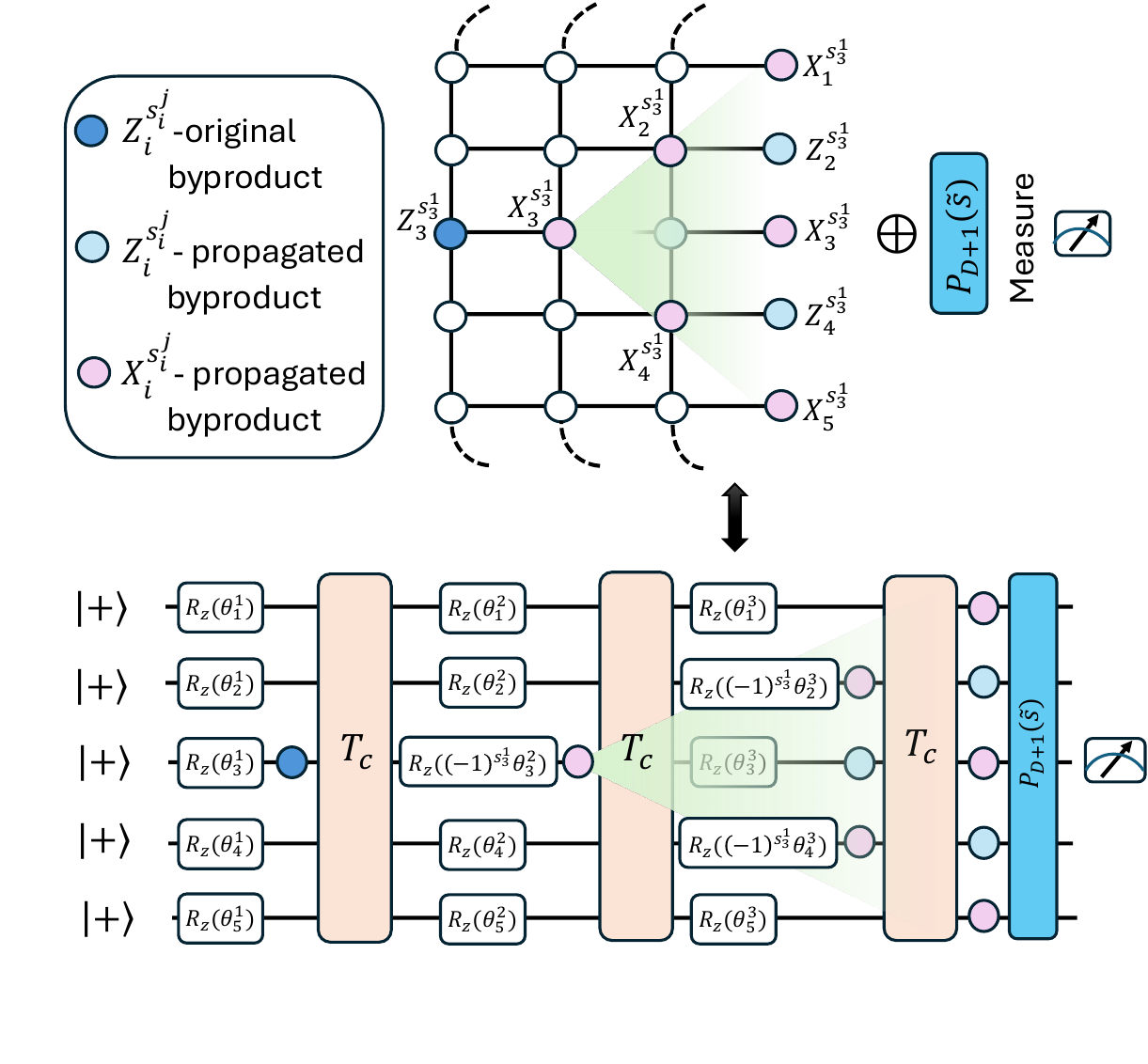}
\caption{\small{\textbf{Propagation of byproducts in $\tilde{\mathcal{E}}_c(\bm{\theta},\bm{p})$ model:}} \textit{Top}: Consider a byproduct $Z_3^{s^1_3}$ on qubit $3$ at position $(3,1)$. Propagating this through the cluster state via repeated stabilizer relations generates $X_i^{s^1_3}$-byproducts on qubits $i$ within the forward light-cone (\textit{green}), which flips (if $s^1_3=1$) the measurement angles of the qubits. For instance, the angle $\theta^2_3$ at position $(3,2)$ is transformed to $(-1)^{s^1_3}\theta^2_3$, following from $X^{s^1_3}_3 R_z(\theta^2_3)\ket{\pm}\bra{\pm}R_z^{\dagger}(\theta^2_3)X^{s^1_3}_3=R_z((-1)^{s^1_3}\theta^2_3)\ket{\pm}\bra{\pm}R_z^{\dagger}((-1)^{s^1_3}\theta^2_3)$ up to a global phase. After propagating through all layers in the cluster, the byproducts accumulate at the final layer (before computational measurements), forming a Pauli string of $X^s$ and $Z^s$ byproducts which depend on outcomes $\bm{s}$. \textit{Bottom}: Equivalent circuit representation of byproduct propagation via repeated Clifford circuits $T_c^{D-j+1}$ ($D=3$, $j\in[D]$) for a byproduct $Z^{s^j_i}_i,~i\in [N]$. In both representations, the accumulated byproducts, at the end of the computation, form a Pauli-string, and the additional $P_{D+1}(\bm{\tilde{s}})$-layer removes this Pauli-string (Clifford part) from the computation and leaves the non-Clifford effect of the byproducts in the intermediate angle flips.} 
\label{fig:byp_prop}
\end{figure}


 As detailed in Ref.~\cite{nautrup2023measurement}, the (universal) family of unitaries, parameterized by $\boldsymbol{\theta}=\{\theta_i^j\}_{i,j}$, that is implementable by the circuit in Fig.~\ref{fig:cqca} (bottom) (assuming no byproducts), is given by

\begin{equation}\label{eq:mbqc_unitary}
    U_c(\bm{\theta})=\prod_{j=D,...,1}\left(T_c\prod_{i=N,...,1}\exp\left(-i\theta^{j}_iZ_i/2\right)\right)
\end{equation}

 which is referred to as the  Clifford Quantum Cellular Automata (CQCA) in Ref.~\cite{nautrup2023measurement}. Eq.~\eqref{eq:mbqc_unitary} can be directly put into correspondence with the circuit in Fig.~\ref{fig:cqca} (without any $Z$-byproduct).

In this way, MBQC on a cluster state naturally gives rise to a PQC \textit{ansatz} for variational quantum algorithms (VQAs) \cite{cerezo2021variational}, as formalized in Eq.~\eqref{eq:mbqc_unitary} (see also \cite{nautrup2023measurement}). 

 An expression similar to Eq.~\eqref{eq:mbqc_unitary}, but including byproducts [see Fig.~\ref{fig:cqca} (bottom-blue)] is (see Ref.~\cite{majumder2024variational})

 \begin{equation}\label{eq:mbqc_unitary_byprod}
    U_c(\bm{\theta}, \bm{s})=\prod_{j=D,...,1}\left(T_c\prod_{i=N,...,1}(Z^{s_i^j}_i)\cdot\exp\left(-i\theta^{j}_iZ_i/2\right)\right)
\end{equation}

where, $\bm{s}\in \{0,1\}^{N\times D}$ is an $N\times D$ matrix with binary elements $s^j_i\in\{0,1\}$, 
representing the measurement outcomes at different sites $(i\in[N],j\in [D])$. It controls the presence ($s^j_i=1$) or absence ($s^j_i=0$) of a byproduct operator on qubit $i$ at layer $j$ as $Z_i^{s_i^j}$ (see Fig.~\ref{fig:cqca}).


Besides using CQCA as the parameterized quantum circuit \textit{ansatz} for the QCBM (Fig.~\ref{fig:gen_l}), another key element, in the model, is the controlled use of these byproducts of $U_c$ in Eq.~\eqref{eq:mbqc_unitary_byprod}. In the context of generative modeling, the authors of Ref.~\cite{majumder2024variational} showed that introducing controlled randomness in the VMBQC model can aid distribution learning. In this approach, instead of correcting byproducts every time during the computation, they assign \emph{trainable correction probabilities} $p_i^j$ for every qubit $i$ at layer $j$ to partially correct those byproducts. To describe the model with these additional trainable correction probabilities, they introduce additional implicit parameters $c_i^{j}\in\{0,1\}$, for all qubits $(i,j)\in [N]\times[D]$, that decides whether to correct ($c_i^{j}=1$) a specific byproduct at position $(i,j)$ or not ($c_i^{j}=0$). The $c_i^{j}$ can then be sampled with probability $p^j_i$ as
\begin{equation}
    c^j_i =
\left\{
	\begin{array}{ll}
		0  & \mbox{with  probability } 1- p^{j}_{i},\\
		1 & \mbox{with probability } p^{j}_{i}
	\end{array}
\right.
\end{equation}
where $p^j_i$ is a  trainable correction probability. 

To incorporate the trainable correction probabilities into the model, the authors in Ref.~\cite{majumder2024variational} replace the measurement outcomes $s_i^j$ with an adaptive correction scheme $\tilde{s}_i^j=(1-c^j_i)\cdot s^j_i$ that depends on the controllable parameters $c_i^j$. When we do not correct a byproduct ($c^j_i=0$, with probability $1-p^j_i$), the probability of a $-1$ outcome ($\ket{-}$ state) is $p(s^j_i=1)=\frac{1}{2}$. Therefore, from the law of total probability, we can write 

\begin{align}\label{eq:sample_prob}
    p(\tilde{s}^j_i = 1) = \frac{1}{2} (1-p^j_i)\nonumber\\
    p(\tilde{s}^j_i = 0) = \frac{1}{2} (1+p^j_i)
\end{align}

Given the set of trainable correction probabilities $\bm{p}=\{p^j_i\}_{i,j}$, and Eq.~\eqref{eq:mbqc_unitary_byprod}, \emph{variational MBQC} gives rise to a family of quantum channels as follows,
\begin{align}\label{eq:mbqc_channel_not_model}
    \mathcal{E}_c(\bm{\theta}, \bm{p})[\rho]=\sum_{\bm{\tilde{s}}\in\{0,1\}^{N\times D}}
    p(\bm{\tilde{s}})~ U_c(\bm{\theta}, \bm{\tilde{s}})~\rho~ U_c^\dagger(\bm{\theta}, \bm{\tilde{s}}).
\end{align}

where $p(\bm{\tilde{s}})=\prod_{i,j}p(\tilde{s}^j_i)=\prod_{i,j}\left(\frac{1+(-1)^{\tilde{s}^j_i}p^j_i}{2}\right)$, with $\bm{\tilde{s}}$ being an $N\times D$ binary matrix with  entries $\tilde{s}^j_i\in \{0,1\}$, and $p^j_i$ is the trainable correction probability for the qubit at $(i,j)$-location. 


In MBQC, measurement-induced byproducts can be propagated through the cluster state by repeatedly applying the stabilizer relations (see Fig.~\ref{fig:byp_prop} \textit{top} with one byproduct $Z^{s^1_3}_3$) which can be understood by Eq.~(1) in Ref.~\cite{majumder2024variational}. In the equivalent circuit representation, this corresponds to commuting the  Pauli-$Z_i^{s^j_i}$ byproducts to the end of the circuit through all the following Clifford circuits $T_c^{D-j+1}$ (for a total depth of $D$), which induces sign flips of the intermediate rotation angles along the way (Fig.~\ref{fig:byp_prop} \textit{bottom} with $D=3$). Once the byproducts reach the end of the circuit, they combine into a single Pauli string (e.g., $\hat{P}=X^{s^{1}_3}_1\otimes Z^{s^{1}_3}_2\otimes X^{s^{1}_3}_3\otimes Z^{s^{1}_3}_4\otimes X^{s^{1}_3}_5$ in Fig.~\ref{fig:byp_prop}). The $P_{D+1}(\bm{\tilde{s}})$-layer in Fig.~\ref{fig:byp_prop} (\textit{bottom}) corresponds to this Pauli byproduct operator $\hat{P}$, as it depends on the measurement outcome $\bm{\tilde{s}}$, and is responsible for removing the accumulated Pauli string $\hat{P}$. In Ref.~\cite{majumder2024variational}, this process is described in more detail, and it serves the purpose of removing the (trivial) Clifford part of the byproducts ($\hat{P}$) while retaining the (non-trivial) non-Clifford effect of the byproducts. This non-Clifford effect of the byproducts influences qubits within a specific forward light cone (\textit{green} in Fig.~\ref{fig:byp_prop}) by inducing sign flips of the corresponding rotation angles (as $X^s R_z(\theta)=R_z((-1)^s\theta)X$). This results in unitary evolution: 

\begin{align}\label{eq:mbqc_unitary_model}
    \tilde{U}_c(\bm{\theta}&, \bm{s})=\\
    P_{D+1}&(\bm{\tilde{s}})\left[\prod_{j=D,...,1}\left(T_c\prod_{i=N,...,1}(Z^{s_i^j}_i)\cdot\exp\left(-i\theta^{j}_iZ_i/2\right)\right)\right]\nonumber
\end{align}

 Similar to Eq.~\eqref{eq:mbqc_channel_not_model}, combining the learnable probabilities $\bm{p} = \{p_i^j\}_{i,j}$ with the above unitary evolution defines a \emph{family of quantum channels }:

\begin{align}\label{eq:mbqc_channel_model}
    \tilde{\mathcal{E}}_c(\bm{\theta}, \bm{p})[\rho]=\sum_{\bm{\tilde{s}}\in\{0,1\}^{N\times D}}
    p(\bm{\tilde{s}})~ \tilde{U}_c(\bm{\theta}, \bm{\tilde{s}})\;\rho \;\tilde{U}_c^\dagger(\bm{\theta}, \bm{\tilde{s}})
\end{align}
where $p(\bm{\tilde{s}})=\prod_{i,j}p(\tilde{s}^j_i)$, and $p(\tilde{s}^j_i)$ is given by Eq.~\eqref{eq:sample_prob}. For further details, see \cite{majumder2024variational, nautrup2023measurement}.

The discussion in this section is intended to summarize the MBQC-to-circuit correspondence and the role of measurement-induced byproducts needed for the present analysis. Detailed derivations of the circuit equivalence, byproduct propagation, and adaptive correction rules can be found in Refs.~\cite{nautrup2023measurement, majumder2024variational}. The results developed in the following sections build on this framework by restricting the correction structure and analyzing the resulting learning performance.

\subsection{Learning models}\label{sec:models}

\begin{figure*}[t]
\centering
\includegraphics[width=0.95\textwidth]{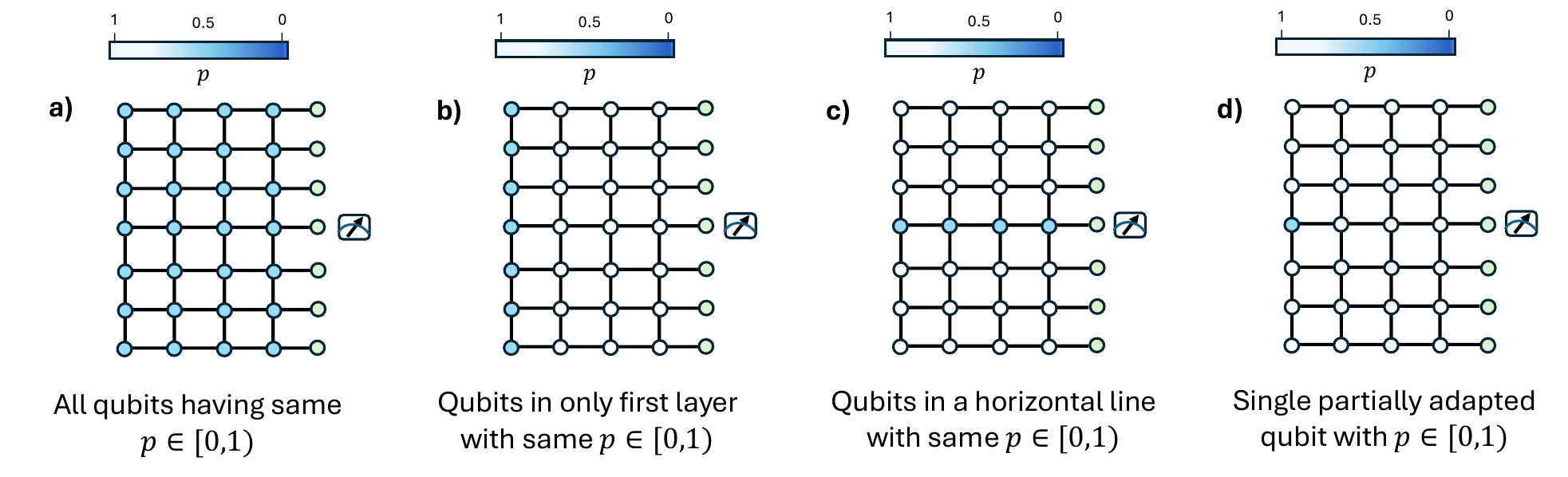}
\caption{\small{\textbf{VMBQC models with one correction probability:}} We consider four variants of the VMBQC model in a restricted setting, where the number of trainable correction probabilities is limited to a single parameter, i.e., $\bm{p}:=\{p^j_i = p\}_{i,j}$.   \textbf{(a)} Channel model $\mathcal{E}_c^{(A)}$ corresponding to  Eq.~\eqref{eq:model_a} where measurement outcomes of all qubits (\textit{blue}) are partially adapted and all are sharing the same correction probability $p<1$. \textbf{(b)}  Channel model $\mathcal{E}_c^{(B)}$ corresponding to  Eq.~\eqref{eq:model_b} where partially corrected qubits are restricted only to the first layer (\textit{blue}), each with the same correction probability $p<1$, while all remaining qubits (\textit{white}), when measured, are always fully corrected ($p=1$). \textbf{(c)} Channel model $\mathcal{E}_c^{(C)}$ corresponding to Eq.~\eqref{eq:model_c} where exactly one partially adapted qubit appears in each layer (\textit{blue}), repeated across all layers (chosen as the central qubit). \textbf{(d)} Channel model  $\mathcal{E}_c^{(D)}$ corresponding to Eq.~\eqref{eq:model_d} has only a single qubit, depicted in \textit{blue}, whose measurement outcome is partially adapted with $p<1$. All the models considered above have periodic boundary conditions, and the qubits in \textit{green} are the output qubits measured in the computational basis.}
\label{Fig:diff_models}
\end{figure*}

In this section, we introduce several simpler variants of the VMBQC model (Eq.~\eqref{eq:mbqc_channel_not_model}). Specifically, we consider models with only one correction probability $\bm{p}:=\{p^j_i=p\}_{i,j}$ for different qubits ($i,j$) in the cluster state. In contrast, in the model from Ref.~\cite{majumder2024variational}, each qubit is assigned a different individually adjustable correction probability. We present four different variants of this kind, depicted in Fig.~\ref{Fig:diff_models}. 


For the model corresponding to Fig.~\ref{Fig:diff_models} $(\mathrm{a})$, one can rewrite Eq.\eqref{eq:mbqc_channel_not_model} as



\begin{equation}\label{eq:model_a}
\begin{aligned}
    \mathcal{E}&^{(A)}_c(\bm{\theta}, \bm{p})[\rho]\\
    &=\sum_{\bm{\tilde{s}}\in\{0,1\}^{N\times D}}\prod_{\substack{j\in[D]\\i\in[N]}}
    \left(\frac{1+(-1)^{\tilde{s}^j_i}p}{2}\right) U_c(\bm{\theta}, \bm{\tilde{s}})~\rho ~U_c^\dagger(\bm{\theta}, \bm{\tilde{s}}).
\end{aligned}
\end{equation}

In Eq.~\eqref{eq:model_a}, all qubits at positions $(i,j)$ for $i\in[N],~j\in [D]$ have the same correction probability $p$. The quantity $\tilde{s}^j_i\in \{0,1\}$, appearing in the exponent, is a binary-valued entry of the matrix $\bm{\tilde{s}}\in \{0,1\}^{N\times D}$ for all $i\in [N], ~j\in [D]$ which specifies whether a byproduct is present $(\tilde{s}^j_i=1)$ at position $(i,j)$ or absent $(\tilde{s}^j_i=0)$ at position $(i,j)$.

To define the model in Fig.~\ref{Fig:diff_models} (b), we consider the matrix 
 
$\tilde{S}^{(B)} :=
\Bigl\{(s_i^j)_{i=1,\dots,N}^{j=1,\dots,D}\in \{0,1\}^{N\times D}
\;\Big|\;  s_i^j = 0 ~\text{if} \ j\neq1
\Bigr\}$, which is a subset of $\{0,1\}^{N\times D}$. Each element $\bm{\tilde{s}} \in \tilde{S}^{(B)}$ is an $N \times D$ matrix whose first $N$ elements are binary numbers, referring only to the qubits (\textit{blue}) in the first layer $j=1$ of the model in Fig.~\ref{Fig:diff_models} (b). The rest of the elements of $\bm{\tilde{s}}\in\{0,1\}^{N\times D}$ are zero, implying the absence of byproducts from other qubits outside the first layer, as those qubits ($i\in[1,N]:=\{1,\cdots ,N\},~j\in[2,D]:=\{2,\cdots,D\}$) are always corrected (\textit{white} qubits in Fig.~\ref{Fig:diff_models} (b)) in this case. 

\begin{equation}\label{eq:model_b}
\begin{aligned}
\mathcal{E}&^{(B)}_c(\bm{\theta}, \bm{p})[\rho]\\
    &=\sum_{\bm{\tilde{s}}\in\tilde{S}^{(B)}}\prod_{\substack{j\in[D]\\i\in[N]}}
    p(\tilde{s}^j_i)~ U_c(\bm{\theta}, \bm{\tilde{s}})~\rho ~U_c^\dagger(\bm{\theta}, \bm{\tilde{s}})
    \\
    &=\sum_{\bm{\tilde{s}}\in\tilde{S}^{(B)}}\prod_{i\in[N]} p(\tilde{s}^1_i)
 \prod_{\substack{j\in[D]\setminus\{1\}\\i\in[N]}} \underbrace{p(\tilde{s}^j_i = 0)}_{=1}
     U_c(\bm{\theta}, \bm{\tilde{s}})~\rho ~U_c^\dagger(\bm{\theta}, \bm{\tilde{s}})
     \\&=\sum_{\tilde{s}\in\tilde{S}^{(B)}} \Bigg(\prod_{i=1}^N p\big(\tilde{s}_i^1\big)\Bigg)\,
U_c(\bm{\theta},\tilde{s})\,\rho\,U_c^\dagger(\bm{\theta},\tilde{s})\\
&=\sum_{\tilde{s}\in\tilde{S}^{(B)}} \Bigg(\prod_{i=1}^N \frac{1+(-1)^{\tilde{s}_i^1}p}{2}\Bigg)\,
U_c(\bm{\theta},\tilde{s})\,\rho\,U_c^\dagger(\bm{\theta},\tilde{s})
\end{aligned}
\end{equation}
where we have used the rule that qubits located at $(i,j)$ for $i\in [N],~ j\in[D]\setminus\{1\}:=\{2,\cdots, D\}$ are always corrected, i.e., $p(\tilde{s}^j_i=0)=1$ for these qubits (Fig.~\ref{Fig:diff_models}~(b)).

For the model in Fig.~\ref{Fig:diff_models} (c), we similarly define $\tilde{\mathop{S}}^{(C)}:=\Bigl\{(s^j_i)_{i=1,\dots, N}^{j=1,\dots, D}\in \{0,1\}^{N\times D} ~\Big|~  s^j_i=0~\text{if}\ ~ i\neq q~ \ \text{for all}~\ j\in [D]\Bigr\}\subset \{0,1\}^{N\times D}$ where any $\bm{\tilde{s}}\in \tilde{\mathop{S}}^{(C)}$ is an $N\times D$ matrix with $D$ binary elements. Each $s^{j}_{q}\in\{0,1\}$ corresponds to the qubit at location $i=q$ in each layer $j\in [D]$ of the model, and the rest of the $(N-1)\times D$ elements are zero, corresponding to the (\textit{white}) qubits that are always corrected. Thus, we can write
\begin{equation}\label{eq:model_c}
\begin{aligned}
    \mathcal{E}&^{(C)}_c(\bm{\theta}, \bm{p})[\rho]\\
    &=\sum_{\bm{\tilde{s}}\in\tilde{S}^{(C)}}\prod_{\substack{j\in[D]\\i\in[N]}}
    p(\tilde{s}^j_i)~ U_c(\bm{\theta}, \bm{\tilde{s}})~\rho ~U_c^\dagger(\bm{\theta}, \bm{\tilde{s}})
    \\&=\sum_{\bm{\tilde{s}}\in\tilde{S}^{(C)}}\prod_{\substack{j=1}}^{D}
    \left(\frac{1+(-1)^{\tilde{s}^j_q}p}{2}\right) U_c(\bm{\theta}, \bm{\tilde{s}})~\rho ~U_c^\dagger(\bm{\theta}, \bm{\tilde{s}})
\end{aligned}
\end{equation}
where we have used $p(\tilde{s}^j_i=0)=1$ for $i\in [N]\setminus\{q\},~ j\in[D]$, i.e.,  for qubits that are always corrected (Fig.~\ref{Fig:diff_models}~(c)). 
As depicted in the figure, in this model, we choose the position ($i=q=4$) for a partially adapted qubit in each layer and repeat the same procedure for all layers $j\in[D]$. 

And lastly, for Fig.~\ref{Fig:diff_models} (d) model, we consider $\tilde{\mathop{S}}^{(D)}:=\Bigl\{(s^j_i)_{i=1,\dots, N]}^{j=1,\dots, D]}\in \{0,1\}^{N\times D}\;\Big|~s^j_i=0~\text{if}~\ (i,j)\neq (q,l)\Bigr\}\subset \{0,1\}^{N\times D}$. Each $\bm{\tilde{s}} \in \tilde{\mathop{S}}^{(D)}$ is a matrix with only one binary element that corresponds to a qubit located at index $(i,j)=(q,l)=(4,1)$, while all other elements are zero. 

Thus, the model can be expressed as
\begin{equation}\label{eq:model_d}
\begin{aligned}
\mathcal{E}&^{(D)}_c(\bm{\theta}, \bm{p})[\rho]\\
    &=\sum_{\bm{\tilde{s}}\in\tilde{S}^{(D)}}\prod_{\substack{j\in[D]\\i\in[N]}}
    p(\tilde{s}^j_i)~ U_c(\bm{\theta}, \bm{\tilde{s}})~\rho ~U_c^\dagger(\bm{\theta}, \bm{\tilde{s}})
    \\&=\sum_{\bm{\tilde{s}}\in\tilde{S}^{(D)}}
    \left(\frac{1+(-1)^{\tilde{s}^l_q}p}{2}\right) U_c(\bm{\theta}, \bm{\tilde{s}})~\rho ~U_c^\dagger(\bm{\theta}, \bm{\tilde{s}})\\
    &=
    \left(\frac{1-p}{2}\right) U_c(\bm{\theta}, \bm{\tilde{s}}_{(q,l)})~\rho ~ U_c^\dagger(\bm{\theta}, \bm{\tilde{s}}_{(q,l)})\\
    &+ \left(\frac{1+p}{2}\right) U_c(\bm{\theta}, \bm{\tilde{s}}=\bm{0})~\rho ~ U_c^\dagger(\bm{\theta}, \bm{\tilde{s}}=\bm{0})
    \end{aligned}
\end{equation}
where we used $p(\tilde{s}^j_i=0)=1$ for $\forall (i,j)\in ([N]\times[D])\setminus \{(q,l)\}$, as the rest of the qubits in the cluster, except the one at $(q,l)$-position, are always corrected (see Fig.~\ref{Fig:diff_models}~(d)). Here $\bm{\tilde{s}}_{(q,l)}\in \{0,1\}^{N\times D}$ is a matrix whose $(q,l)$-th element is $1$, while all other elements are zero, and $\bm{\tilde{s}}=\bm{0}$ is an $N\times D$ matrix with all elements equal to zero. 

Notice that in all the models shown in Fig.~\ref{Fig:diff_models}, there is only one trainable correction probability $p$ that allows for different channel models shown in Eq.~\eqref{eq:model_a}, \eqref{eq:model_b}, \eqref{eq:model_c}, and \eqref{eq:model_d}. In Sec.~\ref{sec:res}, we perform our numerical analysis using these models. 

Throughout this paper, we describe all the models using the cluster-state MBQC representation, as it provides a transparent picture of measurement adaptivity and byproduct propagation. We note that this representation is in one-to-one correspondence with the equivalent CQCA-based circuit description (see Fig.~\ref{fig:cqca}), and both viewpoints are used interchangeably in the analysis of this paper.

\subsection{Loss and gradient}\label{sec:loss}
For training, we employ the Maximum Mean Discrepancy (MMD) \cite{gretton2012k} as an implicit loss function, defined below:

\begin{equation}\label{eq:mmd loss}
\begin{aligned}
    \mathcal{L}(\bm{\theta}, \bm{p})= {} & \mathop{\mathbb{E}}_{\substack{x \sim q_{(\bm{\theta}, \bm{p})}\\ y \sim q_{(\bm{\theta}, \bm{p})}}} [K(x,y)]-2\mathop{\mathbb{E}}_{\substack{x \sim q_{(\bm{\theta}, \bm{p})}\\ y \sim Y}} [K(x,y)] \\
    & +\mathop{\mathbb{E}}_{\substack{x \sim Y\\ y \sim Y}} [K(x,y)]
\end{aligned}
\end{equation}
Here, the $q_{(\bm{\theta}, \bm{p})}$ represents the output distribution of the training model with the set of parameters $\{\bm{\theta}, \bm{p}\}$ (see Fig.~\ref{fig:gen_l}) and $Y(x)$ is some target distribution. The $K(x,y)$ in Eq.~\eqref{eq:mmd loss} is referred to as the kernel function that quantifies the similarity between samples $x$ and $y$. 

The MMD loss function is
particularly suitable for implicit quantum generative models like VMBQC, whereas KL divergence requires access to explicit probability densities and can become ill-defined when the model assigns
zero probability to target events. Moreover, Ref.~\cite{rudolph2024trainability} shows that MMD can avoid the trainability
problems associated with explicit losses such as KL divergence, provided that an appropriate
kernel is chosen.

To train the QCBM, the parameters of the VMBQC model are updated iteratively by computing the gradients of the loss function with respect to the variational angles $\theta^j_i$ and correction probabilities $p^j_i$, i.e., $\frac{\partial \mathcal{L}}{\partial \theta^{j}_{i}}$ and $\frac{\partial \mathcal{L}}{\partial p^{j}_{i}}$, respectively. Refs.~\cite{liu2018d,majumder2024variational} provide more details on how to compute these gradients analytically. The corresponding expression is also provided in Appendix~\ref{app:grad_mmd}. However, in our training models, we have only one variational parameter $p$ shared across different qubits, and thus one can find the gradient of the loss $\frac{\partial \mathcal{L}}{\partial p}$ using the chain rule

\begin{equation}\label{eq:grad_same_p}
    \frac{\partial \mathcal{L}}{\partial p}=\sum_{i,j=1}^{N\times D} \frac{\partial \mathcal{L}(\bm{\theta}, \bm{p})}{\partial p^j_i}\Big{|}_{(\bm{p}:=\{p,\dots,p\})}
\end{equation}

From Eq.~\eqref{eq:grad_same_p}, we observe that the analytical gradient computation requires computing $\frac{\partial \mathcal{L}}{\partial p^{j}_{i}}\Big{|}_{(\bm{p}:=\{p,\dots,p\})}$ at all positions $(i,j)$, which does not speed up the gradient estimation even though it is evaluated at a point determined by the single parameter $p$. However, one can use the finite difference method to estimate the gradient, which requires only one step as follows

\begin{equation}\label{eq:grad_same_p_FD}
    \frac{\partial \mathcal{L}}{\partial p}=\frac{\mathcal{L}(\bm{\theta}, \bm{p}+\bm{\epsilon})-\mathcal{L}(\bm{\theta},\bm{p}-\bm{\epsilon})}{2\epsilon} + \mathcal{O}(\epsilon^2)
\end{equation}
where, $\bm{p}\pm\bm{\epsilon}=\{p^j_i:=(p\pm\epsilon)\}_{i,j}$, for $\epsilon>0$, and $\mathcal{O}(\epsilon^2)$ contain the terms of order $\epsilon^2$ in the Taylor expansion of $\mathcal{L}(\bm{\theta},\bm{p}\pm \bm{\epsilon})$. 

While Eq.~\eqref{eq:grad_same_p} provides the exact gradient of the loss function, it scales poorly with system size. For larger systems, Eq.~\eqref{eq:grad_same_p_FD} is therefore preferred, despite being only an approximation of Eq.~\eqref{eq:grad_same_p}. These gradients are estimated using samples generated by the training and target models. One could also use batch or mini-batch updates that can reduce the statistical variance of the loss and gradient estimates by averaging over multiple data samples and circuit shots before updating the parameters. This may stabilize optimization, including the finite-difference gradient estimator in Eq.~\eqref{eq:grad_same_p_FD}.


\section{Results}\label{sec:res}

In this section, we analyze the expressive power of the single-parameter VMBQC models introduced in Sec.~\ref{sec:models}. We begin by revisiting \textit{Theorem~1} of Ref.~\cite{majumder2024variational} from the perspective of the present work. In particular, we highlight how this result implies that an advantage of the channel-based model over its unitary counterpart can already be achieved by introducing a single additional trainable parameter.

\begin{observation}\label{theorem:toy_example_theorem}
		 Let $N$ be the number of qubits, $D$, the depth of the model, $\bm{\theta}:=\{\theta^j_i|\theta^j_i \in [0,2\pi)\}_{i\in[N],j\in[D]}$, the set of trainable rotation angles, and $\bm{p}:=\{p^j_i=p~|~p\in[0,1)\}_{i\in[N],j=[D]}$, the set of trainable correction probabilities. Then there exist a tuple (N,D,$\bm{\theta}$, $\bm{p}$) for which the set of probability distributions  $\mathcal{Q}_{\mathcal{E}_c}(N,D,\bm{\theta}, \bm{p})$, generated by the $\mathcal{E}_c$ model (in Eq.~\eqref{eq:model_b} and Fig.~\ref{Fig:diff_models}(a)), is strictly larger than the set of distributions $\mathcal{Q}_{U_c}(N,D,\bm{\theta})$, generated by the pure unitary model.
	\end{observation}

This observation is a special case of the result presented in \textit{Theorem 1} of Ref.~\cite{majumder2024variational}, specialized to the model depicted in Fig.~\ref{Fig:diff_models}~(b). The proof follows the same reasoning. In Ref.~\cite{majumder2024variational}, the authors validate the observation \ref{theorem:toy_example_theorem} for the VMBQC model with width $N=3$, depth $D=1$, and for a specific choice of parameters $\bm{\theta}=\{\theta_i^1\}_{i\in[1,3]}$ and $\bm{p}=\{p_i^1=\frac{1}{2}\}_{i\in[1,3]}$. This algebraic result motivates the following numerical investigation into the advantages of the restricted VMBQC model.


In Sec.~\ref{res:channel_learn}, we numerically investigate the performance of the variational quantum channel models in Eq.~\eqref{eq:model_a}-\eqref{eq:model_d} with only a single trainable correction probability by comparing to a fully deterministic (unitary) model given by Eq.~\eqref{eq:mbqc_unitary}. 

In Sec.~\ref{res:lightcone}, we further examine the effect of a single partially corrected qubit (Fig.~\ref{Fig:diff_models}~(d)) on the resulting output distribution of the model. We then extend this analysis to a model with two partially corrected qubits sharing a single correction probability. For this purpose, we focus on the channel model $\tilde{\mathcal{E}}_c(\bm{\theta}, \bm{p})$ defined in Eq.~\eqref{eq:mbqc_channel_model}.



\subsection{Learning quantum channel distributions}\label{res:channel_learn}

\subsubsection{Training process}
First, we generate a dataset with samples drawn from the output distribution of the target model (described in Sec.~\ref{sec:target}), with randomly initialized parameters, and then train different learning models (described in Sec.~\ref{sec:leaning_models}) to learn this target distribution via samples using the MMD loss in Eq.~\eqref{eq:mmd loss}. All learning models are trained for a fixed number of epochs with identical hyperparameters (width and depth of the cluster state), and results are averaged over multiple random initializations, as shown in Fig.~\ref{fig:main_fig}~(a-d). More training details are provided in Appendix~\ref{app:learn_channel}.

\subsubsection{Target model}\label{sec:target}
\begin{figure*}[t]
\centering
\includegraphics[width=1\textwidth]{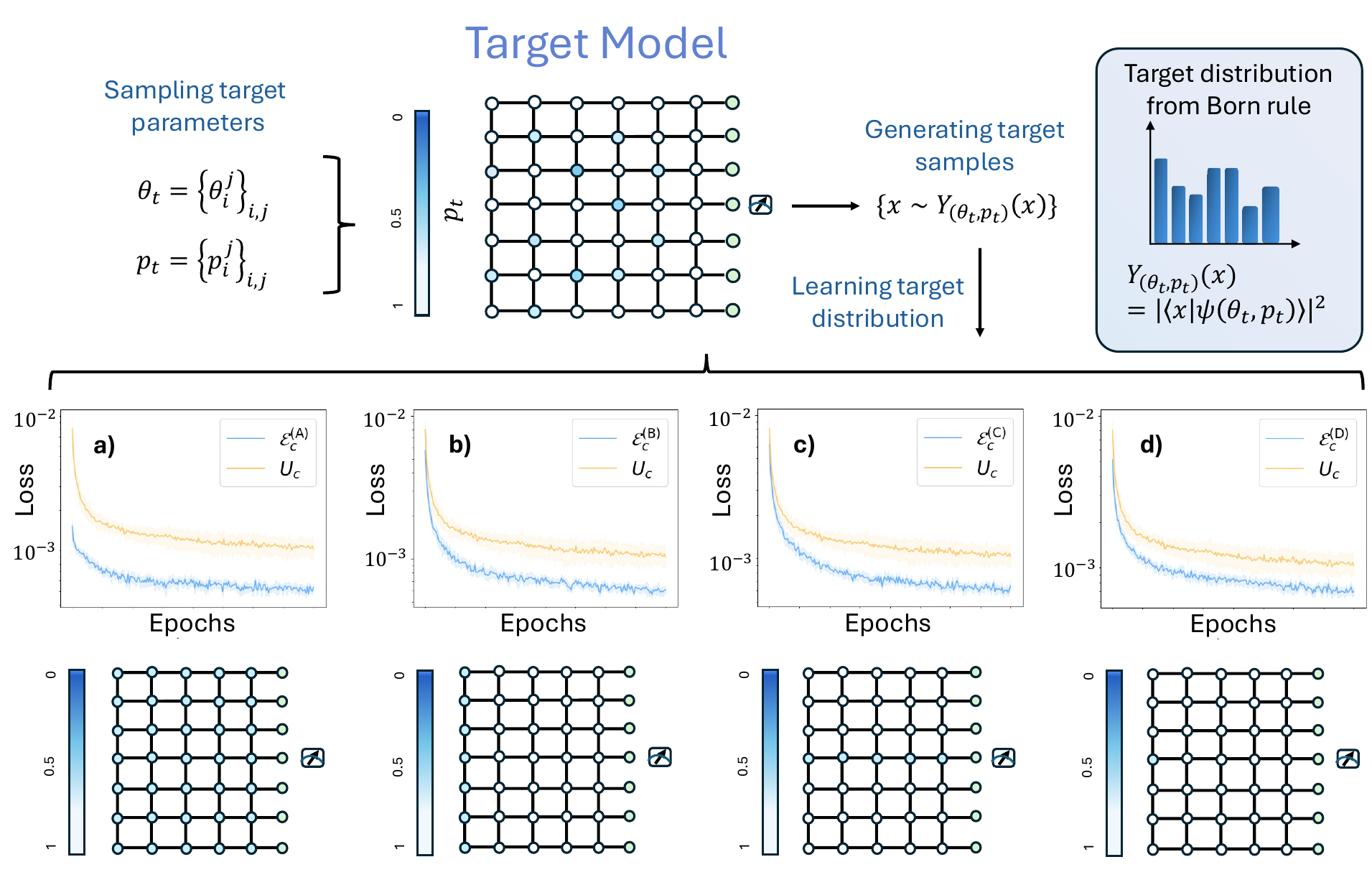}
\caption{\small{\textbf{Learning performances of VMBQC models:}} Four families of channel models, defined in Eqs.~\eqref{eq:model_a}–\eqref{eq:model_d}, together with the unitary model in Eq.~\eqref{eq:mbqc_unitary}, are trained to approximate a target distribution generated by the channel $\mathcal{E}_c(\bm{\theta}_t,\bm{p}_t)$ in Eq.~\eqref{eq:mbqc_channel_not_model}.
\textit{Top row:} The target model has width $N=7$ and depth $D=6$, and generates target samples $x\sim Y_{(\bm{\theta}_t,\bm{p}_t)}(x)$ according to the born rule (in \textit{blue box}). The parameters $\bm{\theta}_t=\{\theta^j_i\in[0,2\pi)\}$ and $\bm{p}_t=\{p^j_i\in[0.9,1)\}$ are drawn independently and uniformly from their specified ranges.
\textit{Middle row} (a-d): Learning performance of the unitary model $U_c$ (yellow) compared to the channel models $\mathcal{E}^{(A)}_c$, $\mathcal{E}^{(B)}_c$, $\mathcal{E}^{(C)}_c$, and $\mathcal{E}^{(D)}_c$ (blue), all with $N=7$ and $D=5$. The qubits (\textit{green}) in the last layer $D=6$ (\textit{bottom row}) correspond to the output qubits. Curves show the mean MMD loss over $200$ epochs, averaged across $10$ random initializations; shaded regions indicate standard deviations.
\textit{Bottom row:} Corresponding channel-based training models, each with a single trainable correction probability $p$, as specified in Eqs.~\eqref{eq:model_a}–\eqref{eq:model_d}. The depicted values of $p$ in the figure are not the optimal values found in our experiments but randomly chosen for illustration. The input to all the learning models is a fixed quantum state $\ket{+}^{\otimes N}$.} 
\label{fig:main_fig}
\end{figure*}

We randomly select an instantiation of the channel model  $\mathcal{E}_c(\bm{\theta}_t,\bm{p}_t)$, defined in Eq.~\eqref{eq:mbqc_channel_not_model} with $\bm{\theta}=\bm{\theta_t}$ and $\bm{p}=\bm{p_t}$, as the target. In particular, for Fig.~\ref{fig:main_fig}, we choose $N=7$ and $D=6$ for the target model. The target is initialized with measurement angles $\bm{\theta}_t:=\{\theta_i^j \in [0,2\pi)\}_{i,j}$ and correction probabilities $\bm{p}_t:=\{p_i^j \in [0.9,1)\}_{i,j}$, both drawn uniformly at random. Note that, in this model, each qubit is assigned its own independent correction probability. The resulting output distribution generated by the target model $\mathcal{E}_c(\bm{\theta}_t,\bm{p}_t)$ serves as the target distribution for the various learning models defined in Eqs.~\eqref{eq:mbqc_unitary}, \eqref{eq:model_a}, \eqref{eq:model_b}, \eqref{eq:model_c}, and \eqref{eq:model_d}.

\subsubsection{Learning models} \label{sec:leaning_models}

The VMBQC learning models considered in this work include a purely unitary model (Eq.~\eqref{eq:mbqc_unitary}) and four families of channel models defined in Eqs.~\eqref{eq:model_a}–\eqref{eq:model_d} (Fig.~\ref{Fig:diff_models} (a-d)), which differ only in the placement of the partially uncorrected qubits on the cluster. We numerically compare the learning performances of these 
channel models with that of the unitary model $U_c(\bm{\theta})$. All channel and unitary models share the same width $N=7$ and depth $D=5$.

\subsubsection{Analysis}\label{sec:analysis_1}
In Fig.~\ref{fig:main_fig} (middle row, panels (a)-(d)), we present the averaged learning curves comparing four families of channel models with the purely unitary model. In all cases, the models are trained to learn the same target distribution generated by another channel model defined in Eq.~\eqref{eq:mbqc_channel_not_model}. Across all panels, a significant separation is observed between the final losses achieved by the unitary model $U_c$ (in \textit{yellow}) and those achieved by the corresponding 
channel models $\mathcal{E}^{(A)}_c$, $\mathcal{E}^{(B)}_c$, $\mathcal{E}^{(C)}_c$, and $\mathcal{E}^{(D)}_c$ (all in \textit{blue}). This consistent performance gap shows that channel-based VMBQC models achieve better learning performance on channel-based target distributions than the unitary VMBQC model, even with only a single additional trainable parameter. 


To assess the performance retained by the restricted parametrization, we additionally compare the single-parameter model $\mathcal{E}^{(A)}_c$ with the full-VMBQC model containing ($N\times D$) independently trainable correction probabilities form Ref.~\cite{majumder2024variational}. The results are presented in Sec.~\ref{app:full_vmbqc_comp}, where all models are trained on the same target distribution and under the same training protocol.

\subsection{Comparison with the full-VMBQC model}\label{app:full_vmbqc_comp} 

\begin{figure}[t] \centering \includegraphics[width=0.38\textwidth] {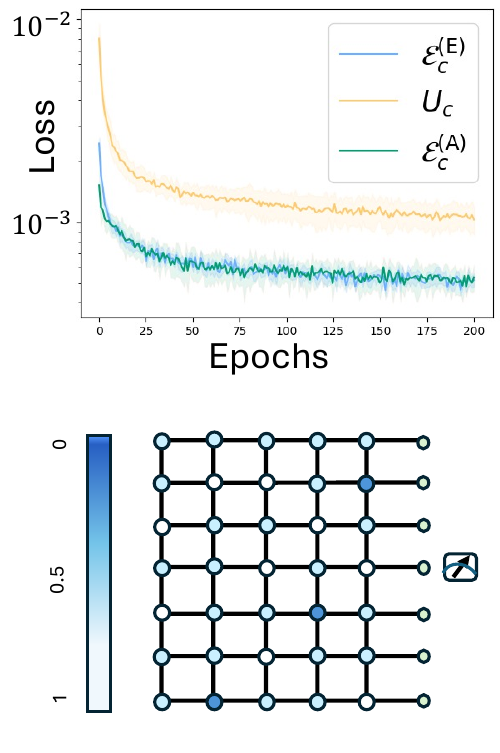} \caption{\small{\textbf{Comparison with the full-VMBQC model.} The unitary model $U_c$, the single-parameter VMBQC model $\mathcal{E}_c^{(\mathrm{A})}$, and the full-VMBQC model $\mathcal{E}_c^{(\mathrm{E})}$ with $N\times D$ independently trainable correction probabilities are trained on the same target distribution using the same training protocol as in Fig.~\ref{fig:main_fig}. The upper panel shows the mean MMD loss over $200$ epochs, averaged over $10$ random initializations, while the shaded regions indicate one standard deviation. The lower panel illustrates the independently trainable correction probabilities $p_i^j$ of the full model across the measured cluster-state qubits. The output qubits are shown in green. The depicted values of $p^j_i$ in the
lower panel of the figure are not the optimal values found in our experiments but randomly chosen for illustration. }} \label{fig:full_vmbqc_comparison} \end{figure}

We compare the unitary model $U_c$, the single-parameter model $\mathcal{E}_c^{(\mathrm{A})}$, and the full-VMBQC channel model, denoted by $\mathcal{E}_c^{(\mathrm{E})}$, with independent correction probabilities $p_i^j$ for every measured qubit $i$ in layer $j$ as defined in Eq.~(10) of Ref.~\cite{majumder2024variational}. All models use the same circuit depth, target distribution, optimizer, and sampling budget as in Fig.~\ref{fig:main_fig}. Note that the full-VMBQC channel model $\mathcal{E}_c^{(\mathrm{E})}$, from Ref.~\cite{majumder2024variational}, belongs to the same family of models as the higher-depth target channel model in Fig.~\ref{fig:main_fig} that generates the target samples. As shown in Fig.~\ref{fig:full_vmbqc_comparison}, the single-parameter model achieves a loss comparable to that of the full model for the target considered, indicating that much of the observed learning performance can be retained despite the substantial reduction in correction-probability parameters. This result suggests that, for certain target distributions, the single-parameter channel model is as expressive as the full-VMBQC model in Ref.~\cite{majumder2024variational}. This does not imply, however, that the single-parameter model can always perform as well as the full-VMBQC model. 

Furthermore, the more spatially restricted single-parameter models converge more slowly and attain higher losses while learning the same target distribution, as shown in Fig.~\ref{fig:main_fig} (b)--(d). Thus, parameter sharing can also reduce learning performance depending on the correction pattern. This comparison is numerical and does not imply equivalence between the two model families.

We note that this comparison also focuses on target correction probabilities $\bm{p}_t:=\{p_i^j \in [0.9,1)\}_{i,j}$, close to the unitary limit. The behavior of the single-parameter model further away from this regime is not systematically studied in
this work.

\subsection{Effect of byproducts on the learning performance}\label{res:lightcone}
So far, we have numerically demonstrated the advantage of using channel-based models with a single correction probability $p$ over unitary models. 
However, it remains unclear in what way this advantage is connected to the position of the byproduct in the cluster. To further analyze and understand the demonstrated advantage, we consider the model in Fig.~\ref{Fig:diff_models}~(d) and compare the advantage between different positions of the byproduct in the 2D grid.
As the model in Fig.~\ref{Fig:diff_models}~(b) contains byproducts in the same time slice, and which are therefore commuting, we can also expect the analysis to carry over to this case.
To highlight the role of a single byproduct, we therefore isolate its effect by considering VMBQC channel models with exactly one partially adapted qubit (see  Fig.~\ref{Fig:diff_models}~(d)), and analyze how the resulting localized measurement-induced byproduct influences the output distribution depending on its position in the cluster state.

Here, we choose the learning model to be the unitary model $U_c$ (Eq.~\ref{eq:mbqc_unitary}), and assess its ability to learn different target models that are based on the channel model in Eq.~\eqref{eq:lightcone_1_main} with one byproduct at a specific position in the cluster. For each target, the unitary learning model is trained with $10$ random instantiations using the MMD loss, and performance is summarized by recording the minimum MMD loss attained by each model instantiation over all training epochs. The resulting distributions of these minimum losses (shown on the $y$-axis) across different targets (shown on the $x$-axis) are presented in Figs.~\ref{fig:lightcone_1} and \ref{fig:lightcone_2} using box plots. 

For the following analysis of the results, we mainly focus on the \textit{lower whisker endpoint (LWE)} (bottom horizontal line) of each distribution, which represents the \textit{minimum non-outlier MMD loss} within that distribution and thus the best observed performance across $10$ learning models, for each target. Furthermore, we refer to each distribution using the color of the box. Further training details are provided in Appendix~\ref{app:eff_byp}.

\subsubsection{Model with one byproduct}\label{sec:1_byproduct}

\begin{figure*}[t]
\centering
\includegraphics[width=0.75\textwidth]{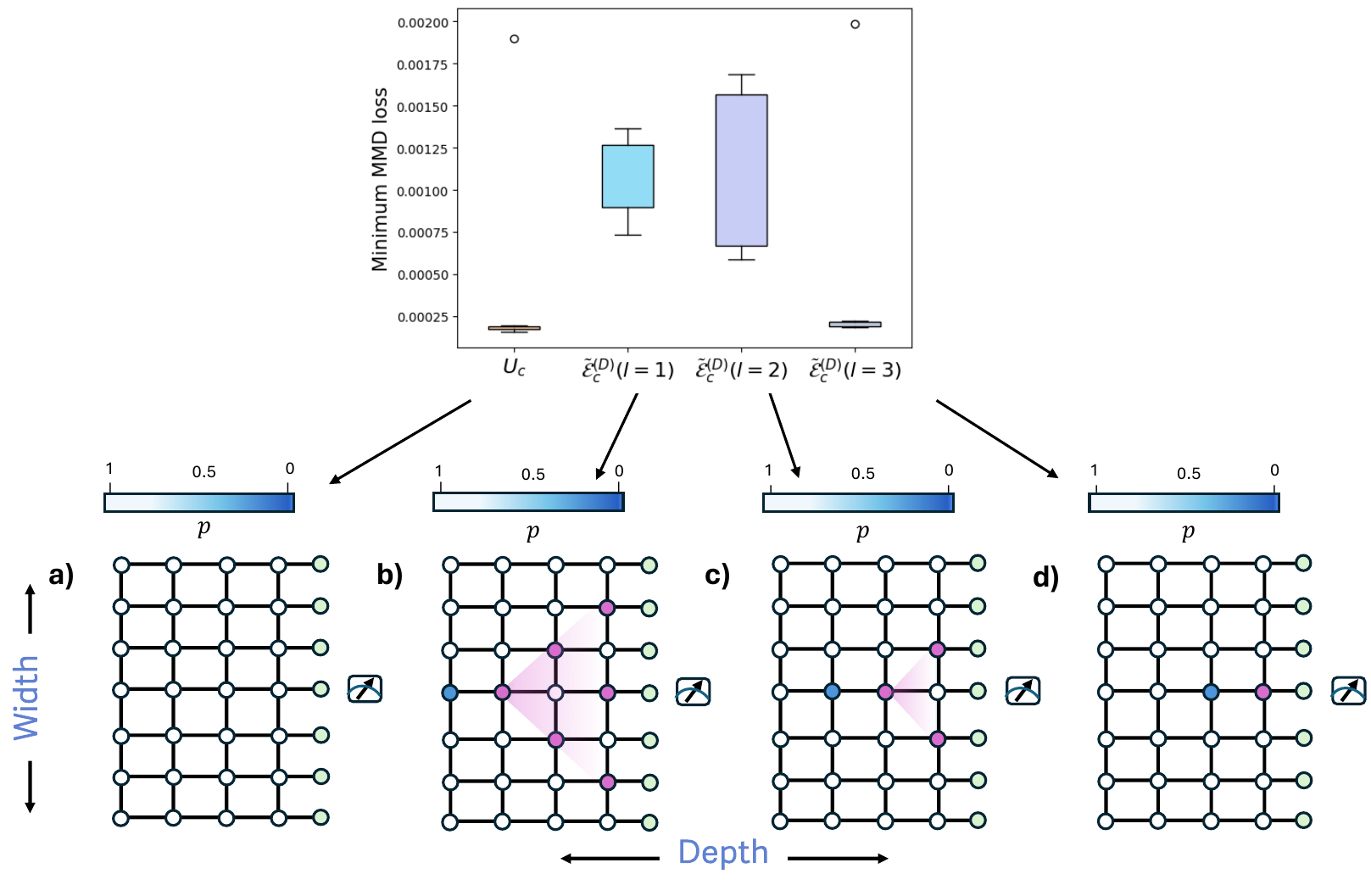}
\caption{\small{\textbf{Effect of a single byproduct:}} Here we examine how well the unitary model $U_c$ can learn the channel model $\tilde{\mathcal{E}}^{(D)}_c(\bm{\theta},\bm{p})$ when only the measurement outcome of a single qubit is probabilistically adapted (see Eq.~\eqref{eq:lightcone_1_main}). The $x$-axis labels the different target models, while the $y$-axis reports the distribution of minimum MMD losses achieved by $10$ independently initialized unitary learning models $U_c$ over $200$ epochs.
\textbf{(a)} \textit{Bottom:} The target distribution is generated by a unitary model with measurement angles $\bm{\theta}=\{\theta^j_i\in[0,1)\}_{i,j}$ drawn uniformly at random. A unitary model is then trained to approximate this distribution. \textit{Top (left-most):} The distribution of \textit{minimum MMD losses} is shown by the box labeled by $U_c$ on the $x$-axis, with a single outlier indicated in \textit{white}. \textbf{(b)} \textit{Bottom:} A single partially uncorrected qubit (in \textit{blue} with $p^l_q=0.135$) is introduced, while keeping $\bm{\theta}$ unchanged, to generate a target distribution. \textit{Top (second from left):} The corresponding MMD distribution is shown with the $\tilde{\mathcal{E}}_c^{(D)}(l=1)$ labeled box.
\textbf{(c)} \textit{Bottom:} The same rotation angles are used, while the partially uncorrected qubit is moved to $(l,q)=(2,4)$ with the same $p^l_q$ as in \textbf{(b)}, yielding the MMD distribution shown by the $\tilde{\mathcal{E}}_c^{(D)}(l=2)$ labeled box.
\textbf{(d)} \textit{Bottom:} Moving the partially uncorrected qubit to $(l,q)=(3,4)$ while keeping the rest the same as in \textbf{(a-c)}. \textit{Top (right-most):} The resulting MMD distribution is shown by the $\tilde{\mathcal{E}}_c^{(D)}(l=3)$ labeled box, with an outlier in \textit{white}.} 
\label{fig:lightcone_1}
\end{figure*}

We now demonstrate the effect of a single byproduct, arising from one partially adapted qubit, on both the output distribution of the VMBQC channel model and the learning performance of a purely unitary model. Specifically, we use the channel model $\tilde{\mathcal{E}}_c(\bm{\theta},\bm{p})$ defined in Eq.~\eqref{eq:mbqc_channel_model} as the target. This model distills the non-Clifford contributions of the byproduct by adding a byproduct-dependent Pauli operator $P_{D+1}(\bm{\tilde{s}})$ that adjusts the final readout basis at the end of the circuit (see Fig.~\ref{fig:byp_prop}). Selecting a specific qubit, characterized by a layer index $l$ and a qubit index $q$, allows Eq.~\eqref{eq:mbqc_channel_model} to be decomposed like Eq.~\eqref{eq:model_d} as follows:


\begin{equation}\label{eq:lightcone_1_main}
\begin{aligned}
\tilde{\mathcal{E}}^{(D)}_c(\bm{\theta}, \bm{p})[\rho]
    &=
    \left(\frac{1-p}{2}\right) \tilde{U}_c(\bm{\theta}, \bm{\tilde{s}}_{(q,l)})\rho \tilde{U}_c^\dagger(\bm{\theta}, \bm{\tilde{s}}_{(q,l)})\\
    &+ \left(\frac{1+p}{2}\right) \tilde{U}_c(\bm{\theta}, \bm{\tilde{s}}=\bm{0})\rho \tilde{U}_c^\dagger(\bm{\theta}, \bm{\tilde{s}}=\bm{0})
    \end{aligned}
\end{equation}
where $\bm{\tilde{s}}_{(q,l)}\in \{0,1\}^{N\times D}$ is a matrix whose $(q,l)$-th element is $1$, while all other elements are zero. Note that in Eq.~\eqref{eq:lightcone_1_main}, $\tilde{U}_c(\bm{\theta}, \bm{\tilde{s}}=\bm{0})=U_c(\bm{\theta})$, i.e., the pure unitary evolution without any byproduct (Eq.~\eqref{eq:mbqc_unitary}). With this setup, we perform a series of simulations in which the unitary model $U_c$, in Eq.~\eqref{eq:mbqc_unitary}, is trained to approximate different target distributions generated by various target models, as shown along the $x$-axis in Fig.~\ref{fig:lightcone_1}. Throughout these simulations, all models, both target and learning, have the same width $N=7$ and depth $D=4$.

We begin with a unitary model $U_c$ (Fig.~\ref{fig:lightcone_1}~(a) with $p=1$) that generates a target distribution for some $\bm{\theta}=\{\theta^j_i\in [0,1]\}_{i,j}$ chosen uniformly at random for all qubits. Since the learning model is also unitary, it is, in principle, capable of learning the target distribution accurately, as reflected by the MMD distribution in \textit{left-most} ($U_c$ labeled on $x$-axis) box (Fig.~\ref{fig:lightcone_1}). However, one training instance converges to a suboptimal local minimum due to the inherent variability in random initialization. Despite this outlier (depicted by a white dot), the key observation from the plot is the minimum loss achieved by the unitary model, demonstrating that, with a suitable parameter initialization, $U_c$ can reach the optimal loss and faithfully reproduce the target distribution.


In Fig.~\ref{fig:lightcone_1}~(b), for the $\tilde{\mathcal{E}}^{(D)}_c(\bm{\theta},\bm{p})$ model, we fix the measurement angles $\bm{\theta}$ to those of the $U_c$ target in (a), and introduce a single partially corrected qubit (in \textit{blue}) with a fixed correction probability $p^l_q=0.135$ (here, $l=1$, $q=4$). This induces a probabilistic byproduct whose forward propagation leads to intermediate angle flips (\textit{dark purple}) within its lightcone (shown in \textit{purple gradient}) (see also Fig.~\ref{fig:byp_prop}). Learning the resulting distribution using unitary models $U_c(\bm{\theta})$, each initialized with different random parameters, yields the MMD loss distribution shown in Fig.~\ref{fig:lightcone_1} (\textit{second from left}, $\tilde{\mathcal{E}}_c^{(D)}(l=1)$ labeled box on $x$-axis), where the LWE is higher compared to the fully unitary target (a). It indicates that even a single probabilistic
byproduct makes the channel distribution harder to learn for the unitary model. The effect is most pronounced for low correction probabilities, typically for $p^l_q \in [0.1, 0.2]$.


In Fig.~\ref{fig:lightcone_1}~(c), we retain the same measurement angles from (a) and move the partially adapted qubit to the next layer, $l=2$, $q=4$. As a result, the LWE of the corresponding channel distribution (Fig.~\ref{fig:lightcone_1}, \textit{third from left}, $\tilde{\mathcal{E}}_c^{(D)}(l=2)$ labeled box) decreases relative to (b) but remains higher than the fully unitary case (a). Shifting the uncorrected qubit reduces the byproduct lightcone size (\textit{purple gradient}), leading to a smaller non-Clifford effect of the byproducts (\textit{dark purple}) and a smaller impact on the output distribution.


In Fig.~\ref{fig:lightcone_1}~(d), we maintain the same setup and move the partially uncorrected qubit to $l=3$, $q=4$. In this configuration, all the unitary learning models reach a very low MMD loss (Fig.~\ref{fig:lightcone_1}, \textit{right-most}, $\tilde{\mathcal{E}}_c^{(D)}(l=3)$ labeled box), with a single outlier due to poor initialization. This illustrates that while a single probabilistic byproduct can make the output distribution of the channel model harder to learn for the unitary model, its impact depends strongly on its position. Notably, placing the uncorrected qubit in the final layer ($l=4$) has no effect, as no further angle flips occur and the output remains effectively unitary.

\subsubsection{Model with two byproducts }\label{sec:2_byproduct}

\begin{figure*}[t]
\centering
\includegraphics[width=0.75\textwidth]{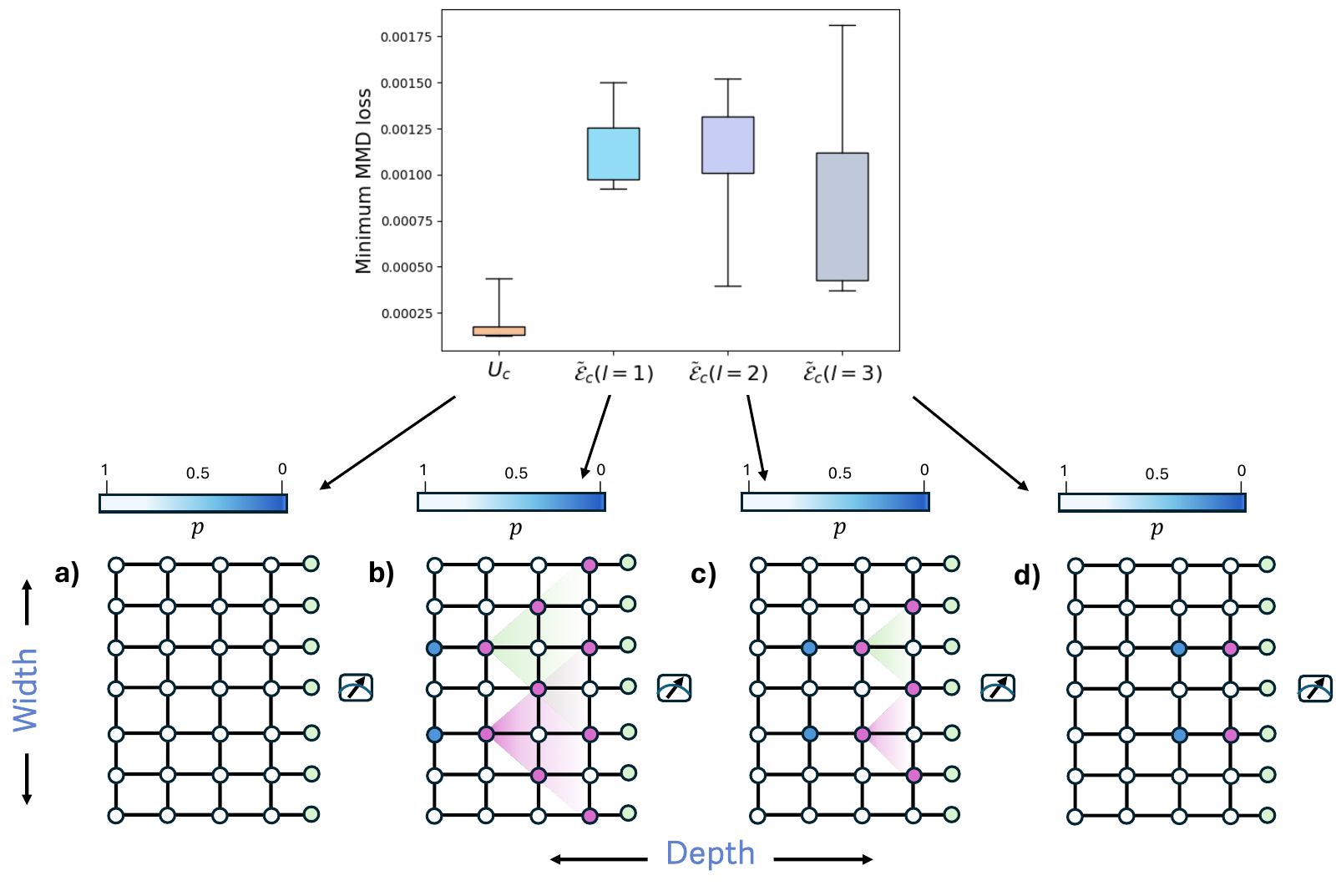}
\caption{\small{\textbf{Effect of two byproducts:}} Here we examine how well the unitary model $U_c$ can learn the channel model $\tilde{\mathcal{E}}_c(\bm{\theta},\bm{p})$ with two partially uncorrected qubits whose measurement outcomes are probabilistically adapted.  The $x$-axis labels the different target models, while the $y$-axis reports the distribution of minimum MMD losses achieved by $10$ independent randomly initialized unitary models $U_c$ over $200$ epochs. \textbf{(a)} \textit{Bottom:} The target distribution is generated from the unitary model itself with $\bm{\theta}=\{\theta^j_i\in[0,1)\}_{i,j}$ drawn uniformly randomly. \textit{Top (left-most):} The resulting minimum MMD distribution is shown with the box labeled by $U_c$ on the $x$-axis. \textbf{(b)} \textit{Bottom:} We choose two uncorrected qubits (in \textit{blue}) for some fixed $p^1_3=p^1_5=0.1$ for both of them, keeping the $\theta$ same as earlier. \textit{Top (second from left):} The minimum MMD losses, for the corresponding output distribution, are shown using the $\tilde{\mathcal{E}}_c(l=1)$ labeled box. \textbf{(c)} \textit{Bottom:} The uncorrected qubits are moved to $~l=2,~q=3,5$, with the same $p^l_q$ and $\bm{\theta}$ as before. \textit{Top (Third from left):} The corresponding MMD distribution, for this target model, is shown by the $\tilde{\mathcal{E}}_c(l=2)$ labeled box. \textbf{(d)} \textit{Bottom:} The uncorrected qubits are placed in layer $l=3$ and the resulting MMD distribution is shown with the corresponding $\tilde{\mathcal{E}}_c(l=3)$ labeled box.} 
\label{fig:lightcone_2}
\end{figure*}

In this section, we extend the previous analysis with byproducts at two different positions but with same correction probability and their impact on the output distribution of the $\tilde{\mathcal{E}}_c(\bm{\theta},\bm{p})$ model. Similar to Sec.~\ref{sec:1_byproduct}, we conduct a series of simulations with four different target models (Fig.~\ref{fig:lightcone_2} (a-d)), and the results follow a trend similar to that in Fig.~\ref{fig:lightcone_1}. All target models and the unitary learning model have the same width $N=7$, and depth $D=4$, with the last layer being the measurement layer.

We start again with a unitary target model $U_c$ (Fig.~\ref{fig:lightcone_2}~(a)), which defines a target distribution using measurement angles $\bm{\theta}=\{\theta^j_i\in [0,1]\}_{i,j}$ drawn uniformly at random (distinct from those in Sec.~\ref{sec:1_byproduct}). Since the learning model is also unitary, it successfully reproduces the target distribution, as indicated by the low MMD losses (\textit{left-most}, $U_c$ labeled box on $x$-axis).

In Fig.~\ref{fig:lightcone_2}~(b), we keep the same measurement angles $\bm{\theta}$ used in (a) and introduce two partially uncorrected qubits (highlighted in \textit{blue}) that share a single fixed correction probability $p_3^1=p_5^1=0.1$. The corresponding byproduct lightcones (\textit{green} and \textit{purple} gradients) overlap, leading to angle flips (shown in \textit{dark purple}) within their combined region. Training the unitary models, with random initializations, on the resulting target distribution yields a distribution of minimum MMD losses, shown in Fig.~\ref{fig:lightcone_2} (\textit{second from left}, $\tilde{\mathcal{E}}_c(l=1)$ labeled box on $x$-axis). As expected, the LWE of the MMD loss distribution increases compared to the purely unitary target, with the effect being most pronounced for small values of $p_q^l$, typically within $ [0.1,0.2]$.

In Fig.~\ref{fig:lightcone_2}~(c), both partially adapted qubits are shifted to the next layer ($l=2$ with $q=3,5$), while keeping the same measurement angles $\bm{\theta}$ as earlier. The distribution of MMD (Fig.~\ref{fig:lightcone_2}~ \textit{third from left}, $\tilde{\mathcal{E}}_c(l=2)$ labeled box) shows that the LWE decreases relative to (b) but remains higher than the pure unitary case (a). The placement of the byproducts shrinks the combined lightcone (green and purple gradients) compared to (b), resulting in a smaller non-Clifford effect of the byproducts (shown in \textit{dark purple}) and lower loss (LWE) than (b).

Lastly, in Fig.~\ref{fig:lightcone_2}~(d), the partially uncorrected qubits are placed at layer $l=3$ while the rest remain the same as earlier. The corresponding distribution of the MMD losses is shown using the \textit{right-most} ( $\tilde{\mathcal{E}}_c(l=3)$ labeled) box in Fig.~\ref{fig:lightcone_1}. However, notice that the LWE remains higher than that of the fully unitary target (a). 

In this section, we have been able to partially attribute the learning advantage demonstrated in the previous section (Sec.~\ref{res:channel_learn}) to the position of the byproducts in the cluster. The results in Secs.~\ref{sec:1_byproduct} and \ref{sec:2_byproduct} highlight that, while randomness in the form of probabilistically adapted qubits can influence the output distribution, its impact depends strongly on their position in the cluster state.

Since the models considered in Secs.~\ref{sec:1_byproduct} and \ref{sec:2_byproduct}, as well as the other models used in the numerical analysis, are translationally invariant and use periodic boundary
conditions, the absolute position of a measurement-induced byproduct within a given layer has the same light-cone effect on the rest of the qubits in the cluster state as shown in Fig.~\ref{fig:lightcone_1}. Therefore, the position of the byproduct does
not affect the output distribution significantly. The relevant quantity is instead its distance from the final
readout layer. In Fig.~\ref{fig:lightcone_1} and \ref{fig:lightcone_2}, we observe that larger distances generally influence the output
distributions more significantly.

\section{Conclusion and outlook}\label{conclusion}
In this work, we investigated the variational measurement-based quantum computation (VMBQC) framework for generative modeling introduced in Ref.~\cite{majumder2024variational}. VMBQC leverages the intrinsic indeterminacy of measurement outcomes in MBQC to realize quantum channels as learning models. Specifically, Ref.~\cite{majumder2024variational} introduced independent trainable parameters for each qubit in the MBQC that govern whether or not random measurement outcomes in MBQC are corrected. In contrast, the present work reduces these additional classical resources, namely the trainable correction probabilities, to a single shared parameter while preserving an expressive advantage over the corresponding purely unitary model. This setting substantially reduces the sampling overhead associated with gradient estimation, since only one additional parameter must be differentiated, compared with the full-VMBQC channel model of Ref.~\cite{majumder2024variational}, which requires estimating gradients with respect to $N\times D$ additional parameters, where $N$ is the height of the cluster state and $D$ is its depth.

We demonstrate the effectiveness of our \textit{simplified} channel VMBQC approach, both algebraically and numerically, through various learning tasks, benchmarking it against purely unitary models. Our results show that channel-based models, even when endowed with only a single trainable correction probability, can generate output distributions that cannot be learned, up to a desired accuracy, by the corresponding unitary models. Importantly, these numerical results are robust: the qualitative behavior and conclusions remain unchanged under different random choices of (target) model parameters within the same range, indicating that the observed advantage is not an artifact of specific instances but reflects an intrinsic property of the model class. For the benchmark considered, the single-parameter model retains much of the task-relevant expressive capability of the full-VMBQC model and achieves comparable learning performance. However, we do not establish a general quantitative characterization of how much expressive power is retained by the restricted VMBQC channel model family compared to the model family with independent correction probabilities as introduced in Ref.~\cite{majumder2024variational}, across arbitrary target distributions.

We also point out that the success of these learning models might depend on the underlying target distribution. Furthermore, properly chosen hyperparameters, particularly the choice and initialization of the shared correction probability, can affect the performance of the model. Most importantly, the benefit of introducing such measurement-induced classical randomness is expected to be most pronounced in the shallow-circuit regime, since sufficiently deep unitary models may already access a much larger portion of the relevant state space.

Moreover, we show that the placement of the random byproducts plays a critical role in determining their impact on the output distribution. By strategically selecting which qubits retain byproducts, one can utilize the channel characteristics of the VMBQC model while keeping the circuit width and depth fixed and the classical control resources minimal. Together, these results provide concrete design principles for quantum generative models that exploit measurement-induced randomness with very few classical control resources within the MBQC framework. 


\section{Data availability}

The code necessary to reproduce the results presented in this paper is available at \cite{Majumder_VMBQC_1p_2026}.

\section{Acknowledgements}
The authors thank Marius Krumm for useful discussions. This research was funded in part by the Austrian Science
Fund (FWF) [SFB BeyondC F7102, DOI: 10.55776/F71;
WIT9503323, DOI: 10.55776/WIT9503323]. For open ac-
cess purposes, the author has applied a CC BY public copy-
right license to any author accepted manuscript version
arising from this submission. This work was also supported
by the European Union (ERC Advanced Grant, QuantAI,
No. 101055129). The views and opinions expressed in
this article are however those of the author(s) only and do
not necessarily reflect those of the European Union or the
European Research Council - neither the European Union
nor the granting authority can be held responsible for them.

\bibliography{ref}

\appendix
\section{Gradients of loss function}\label{app:grad_mmd}

For training purposes, we require calculating the gradient of the loss function in Eq.~\ref{eq:mmd loss}. The resulting gradient with respect to a variational  measurement angle $\theta^j_i$ (see Ref.~\cite{liu2018d}) of the VMBQC model is
\begin{equation}\label{app:grad theta}
\begin{aligned}
    \frac{\partial \mathcal{L}}{\partial \theta^{j}_{i}} = {} & \mathop{\mathbb{E}}_{\substack{x \sim q_{(\bm{\theta^{+}},\bm{p})}\\ y \sim q_{\phi}}} [K(x,y)]-\mathop{\mathbb{E}}_{\substack{x \sim q_{(\bm{\theta^{-}},\bm{p})}\\ y \sim q_{\phi}}} [K(x,y)] \\
      & -\mathop{\mathbb{E}}_{\substack{x \sim q_{(\bm{\theta^{+}},\bm{p})}\\ y \sim Y}} [K(x,y)]+\mathop{\mathbb{E}}_{\substack{x \sim q_{(\bm{\theta^{-}},\bm{p})}\\ y \sim Y}} [K(x,y)]
\end{aligned}
\end{equation}

Here, $q_{(\bm{\theta^{+}}, \bm{p})}$ and $q_{(\bm{\theta^{-}}, \bm{p})}$ are the output distributions of the VMBQC channel model with parameters $\bm{\theta^{\pm}} = \bm{\theta} \pm \frac{\pi}{2} \bm{e} ^{j}_{i}$ and $\bm{e}^{j}_{i}$ is a vector with a $1$ at position $(i,j)$ and $0$s elsewhere. The parameter shifting is done for each angle individually, while the other angles remain unchanged. The $Y$ is the target distribution generated from the target model.
\vspace{8pt}

In the channel models considered for training in this study, there is a single trainable correction probability $p^j_i=p$ shared across different qubits $(i,j)$. Thus, we need to compute the gradient only with respect to one $p^j_i$, for some $(i,j)$. In Ref.~\cite{majumder2024variational}, the general form of the gradient with respect to any $p^j_i$ is provided as 
\begin{equation}\label{app:gradp}
\begin{aligned}
    \frac{\partial \mathcal{L}}{\partial p^{j}_{i}} = {} & 2(\mathop{\mathbb{E}}_{\substack{x \sim q_{\bm{\phi}^{1}}\\ y \sim q_{\bm{\phi}}}} [K(x,y)]-\mathop{\mathbb{E}}_{\substack{x \sim q_{\bm{\phi}^{0}}\\ y \sim q_{\bm{\phi}}}} [K(x,y)]) \\
  & -2(\mathop{\mathbb{E}}_{\substack{x \sim q_{\bm{\phi}^{1}}\\ y \sim Y}} [K(x,y)]-\mathop{\mathbb{E}}_{\substack{x \sim q_{\bm{\phi}^{0}}\\ y \sim Y}} [K(x,y)])
\end{aligned}
\end{equation}

Here, $\bm{\phi}$ is the set of unchanged parameters, i.e., $\bm{\phi}=(\bm{\theta},\bm{p})$, and $\bm{\phi}^{1}=(\bm{\theta},\bm{p}^{1})$ and $\bm{\phi}^{0}=(\bm{\theta},\bm{p}^{0})$, where 
\begin{equation}
    (p^{b})^{j'}_{i'} :=
    \begin{cases}
    p^{j'}_{i'} & \text{ for } i',j' \neq i,j\\
    b & \text{ for } i',j' = i,j
    \end{cases}
\end{equation}

for the $i'$-th qubit and $j'$-th layer, and $b\in\{0,1\}$. While calculating the gradient of the loss $\mathcal{L}$ with respect to $p_i^j$, the other correction probabilities $p^{j'}_{i'},~ (i^{'},j^{'})\neq (i,j)$, remain unchanged whereas the only modification occurs at the $(i,j)$-th location, where the correction probability $p^j_i$ is set to $0$ for $b=0$ and $1$ for $b=1$. The detailed derivation of this gradient is provided in \cite{majumder2024variational}.
To find the gradient with respect to any other $p^{j'}_{i'}=p,~ (i'j')\neq (i,j)$, we use the chain rule
\begin{equation}\label{app:grad_same_p}
    \frac{\partial \mathcal{L}}{\partial p}=\sum_{i,j=1}^{N\times D} \frac{\partial \mathcal{L}(\bm{\theta}, \bm{p})}{\partial p^j_i}\Big{|}_{(\bm{p}:=\{p,\dots,p\})}.
\end{equation}


\section{Training details}
All the training in this paper is performed with the Adagrad~\cite{duchi2011adaptive} optimizer, using the gradients in Eqs.~\ref{app:grad theta} and~\ref{app:gradp}.

\subsection{Learning quantum channel distribution}\label{app:learn_channel}
In this appendix, we provide the training details corresponding to the results presented in Sec.~\ref{res:channel_learn}. We employ gradient-based optimization to train the learning models, using the gradients derived in Eqs.~\eqref{app:grad theta} and \eqref{app:gradp}. Specifically, both the variational measurement angles $\bm{\theta}$ and the correction probabilities $\bm{p}$ are updated during training. Distinct learning rates are used for these parameters: a fixed learning rate of $0.1$ for the measurement angles $\bm{\theta}$, and $0.2$ for the correction probability $p^j_i=p$ shared across different qubits $(i,j)$. A systematic hyperparameter optimization is beyond the scope of this manuscript. We additionally employ the Adagrad optimizer, which adaptively rescales the effective learning rates during training based on the accumulated gradient history.

 For each experiment, we first generate a dataset of $8000$ samples from a target distribution defined by a VMBQC channel model. The target model parameters, measurement angles $\bm{\theta}_t =\{\theta^j_i\in [0,2\pi)\}_{i.j}$ and correction probabilities $\bm{p}_t=\{p^j_i=p \in [0.9,1)\}_{i,j}$, are drawn independently and uniformly at random. Both unitary and channel-based VMBQC models are then trained to learn this target distribution.

 All learning models initialize their variational measurement angles $\bm{\theta}$ uniformly at random from $[0,2\pi)$. For the channel models, the single trainable correction probability $p$ is initialized uniformly from the interval $[0.85,1)$. 
 
We choose the correction probabilities in the target model of Fig.~\ref{fig:main_fig} to be $\bm p_t=\{p^j_i\in [0.9,1)\}$, i.e., close to the unitary regime so that the target distribution remains sufficiently nonuniform~\cite{majumder2024variational} and the comparison probes a nontrivial learning task. Otherwise, both models could approximate an almost uniform target distribution without revealing a meaningful separation. 
 
Initializing the correction probability $p$ for the learning model close to unity (between $[0.85,1)$) is empirically favorable, as smaller values tend to produce overly uniform output distributions~\cite{majumder2024variational}, from which optimization is empirically less effective.

 During training, each model generates $8000$ samples per epoch, matching the size of the target dataset. These samples, together with the target samples, are used to evaluate the MMD loss function (Eq.~\eqref{eq:mmd loss}) on a classical computer (see Fig.~\ref{fig:gen_l}). Model parameters are then updated after each epoch using the gradients given in Eqs.~\eqref{app:grad theta} and \eqref{app:gradp}. Each model is trained for a total of $200$ epochs.

 To ensure robustness and statistical reliability, all experiments are repeated with $10$ independent random initializations of the model parameters. The reported results correspond to the average performance across these runs, as shown in Fig.~\ref{fig:main_fig}~(a-d). All learning models in this section are trained using identical hyperparameters, including the cluster-state width $N$, depth $D$, and dataset size.

\subsection{Effect of byproducts on the learning performance}\label{app:eff_byp}
This appendix provides the training details corresponding to the results presented in Secs.~\ref{sec:1_byproduct} and \ref{sec:2_byproduct}. In contrast to Sec.~\ref{res:channel_learn}, here we fix the learning model to be the unitary VMBQC $U_c$ in Eq.~\eqref{eq:mbqc_unitary} and evaluate its ability to learn a variety of target models.

For each target model, the parameters, measurement angles $\bm{\theta}=\{\theta^j_i\in [0,1]\}_{i,j}$ and correction probabilities $\bm{p}=\{p^j_i=p\in[0.1,0.2]\}_{i,j}$ are sampled randomly, and a dataset of $8000$ samples is generated from the corresponding output distribution. To learn each target distribution, we train $10$ independent instances of the unitary model $U_c$, each initialized with different random parameters. Training is performed using the MMD loss function (Eq.~\eqref{eq:mmd loss}), which is evaluated at every epoch and used to update the model parameters.

Each unitary learning model is trained for $200$ epochs. While the loss is monitored throughout training, performance is summarized by recording the minimum MMD loss achieved by each model across all $200$ epochs. The distributions of these minimum losses are shown as box plots in Figs.~\ref{fig:lightcone_1} and \ref{fig:lightcone_2}, where each box (and the vertical line) corresponds to the distribution of $10$ minimum MMD losses obtained by the $U_c$ model, for a given target model. Across all experiments in this section, we use a fixed learning rate of $0.1$ for the unitary learning model.

The figures in these subsections employ standard box plots to summarize the distributions of MMD losses across different target models. The $x$-axis labels the different target models under consideration, while the $y$-axis reports the corresponding distribution of minimum MMD loss values. Each box spans the interquartile range (IQR), representing the central $50\%$ of the data. The whiskers (vertical lines), containing the rest of the data, extend to the most extreme data points within $1.5\times$IQR from the quartiles, with values beyond this range treated as outliers. Throughout our analysis, we focus on the lower whisker endpoint (bottom horizontal line), which corresponds to the minimum non-outlier MMD loss achieved for a given target model and learning method, and thus reflects the best performance without reliance on statistical outliers.







\end{document}